\documentclass[twocolumn,superscriptaddress,amsmath,amssymb,aps,prl]{revtex4-1}
\usepackage{graphicx, float}
\usepackage[columnwise]{lineno}
\usepackage{xcolor}
\usepackage[normalem]{ulem}

\newcommand{\rme}{\mathrm{e}}

\begin{document}

\title{Suppressed spontaneous emission for coherent momentum transfer}

\author{Xueping Long}
\email{gregllong@gmail.com}
\affiliation{University of California Los Angeles}
\author{Scarlett S. Yu}
\affiliation{University of California Los Angeles}
\author{Andrew M. Jayich}
\affiliation{University of California Santa Barbara}
\author{Wesley C. Campbell}
\affiliation{University of California Los Angeles}

\date{\today}

\begin{abstract}
Strong optical forces with minimal spontaneous emission are desired for molecular deceleration and atom interferometry applications.
We report experimental benchmarking of such a stimulated optical force driven by ultrafast laser pulses. We apply this technique to accelerate atoms, demonstrating up to an average of $19$ $\hbar k$ momentum transfers per spontaneous emission event. This represents more than an order of magnitude improvement in suppression of spontaneous emission compared to radiative scattering forces.  For molecular beam slowing, this technique is capable of delivering a many-fold increase in the achievable time-averaged force to significantly reduce both the slowing distance and detrimental losses to dark vibrational states. 
\end{abstract}

\maketitle

The directed, narrow-band light emitted by lasers has been used to great effect to manipulate the motion of gas-phase atoms, leading to a diverse set of applications \cite{Anderson198BoseEinsteinCondensate,Ludlow2015Clock,Cronin2009interferometer,Haffner2008Quantum}.
In contrast to atoms, the rich internal structures of polar molecules and their readily available long-range and anisotropic dipolar interactions make ultracold molecules uniquely promising candidates for precision measurements \cite{Wilkening1984PTviolation,Hinds1997testingTimeReversalSymmetry,Vutha2010EDMTiO,Altunta2018NuclearSpinParityViolation,Kozyryev2017PrecisionMeasurement}, quantum information processing \cite{Demille2002QuantumComputing,Andre2006QIinterface,Rabl2006QuantumMemory,Hudson2018QuantumLogic,Baranov2012QuantumSimulation,Micheli2006QuantumSimulation} and quantum chemistry \cite{Ospelkaus2010QuantumChemistry,Krems2008ColdControlledChemistry}.  However, for molecules, spontaneous emission populates excited
vibrational states, which has largely precluded the adaptation of atomic laser cooling techniques for molecules.

Recently, the workhorse of ultracold atomic physics, the magneto-optical trap (MOT), has been successfully demonstrated with some carefully chosen diatomic molecules \cite{Barry2014MOTofDiatomicMolecule,Norrgard2016RadioFrequencyMOT,Steinecker2016ImprovedRadioFreqMOT,Truppe2017SubDopplerMOT,Anderegg2017HighDensityRadioFreqMOT,Anderegg2018LaserCoolingMOT}. Despite this substantial step forward, the largest number of molecules that have been trapped in a MOT ($\approx 10^5$ \cite{Anderegg2017HighDensityRadioFreqMOT}) is still orders of magnitude less than a typical atomic MOT, limited by the small fraction of molecules that can be slowed from a molecular beam to the MOT capture speed \cite{McCarron2018TowardsSlowMolecules}. Further, extension of this technique to molecules with higher vibrational branching probability (such as polyatomics) will likely require new methods for beam deceleration.

While the most commonly used laser deceleration methods employ spontaneous radiation pressure, the time-averaged force is limited to a low value by the need to wait for spontaneous decay after each $\hbar k$ of momentum transfer.  For molecules, slowing via spontaneous scattering has been limited to a handful of specially-chosen diatomic species \cite{Truppe2017ChirpSlowing,Barry2012RadiationPressureSlowing, Zhelyazkova2014CaFCoolingSlowing, Hemmerling2016CaFWhiteLightSlowing, Yeo2015RotationalStateMicrowaveMixing,Petzold2018ZeemanSlower} with extremely low vibrational branching probabilities \cite{DiRosa2004Laser}.  Moreover, multiple molecular transitions must be driven that connect various ground states to the same excited state, which further reduces the time-averaged force that can be achieved \cite{Berkeland2002Destabilization}.  As a result, radiative deceleration of molecular beams leads to long slowing lengths and low trap capture efficiencies associated with molecule loss from transverse velocity spread and spontaneously populated excited vibrational states.

For atom interferometry \cite{Johnson2015Sensing,Campbell2017Rotation} (including fast entangling gates with trapped ions \cite{GarciaRipoll2003Speed,Duan2004Scaling,WongCampos2017TwoIonGate}), coherent forces are needed to manipulate phase space separation.  In these cases, even a single spontaneously emitted photon can carry ``which way" information that will decohere the superposition, entirely precluding the use of spontaneous radiation pressure for these applications.  Further, strong forces are desired to effect large separation in a short interaction time, and coherent, spin-dependent momentum kicks \cite{Mizrahi2013Ultrafast} are particularly attractive \cite{Jaffe2018Efficient,Bentley2015IonGate,Heinrich2019Ultrafast}.

To address these needs, various optical forces that utilize \emph{stimulated} emission are being pursued.  For stimulated forces, a reasonable figure of merit for evaluating the gain in requisite cycle closure of stimulated over spontaneous scattering is the average momentum transferred (in units of the photon momentum, $\hbar k$) per spontaneous emission event, which we denote by the symbol $\Upsilon$.  For spontaneous scattering, $\Upsilon=1$.  For most stimulated scattering schemes, the stimulated processes can be driven quickly compared to the spontaneous emission lifetime, and the stimulated force can therefore be stronger than the spontaneous scattering force by a factor of approximately $\Upsilon$.

\begin{figure}
    \centering
    \includegraphics[scale=0.25]{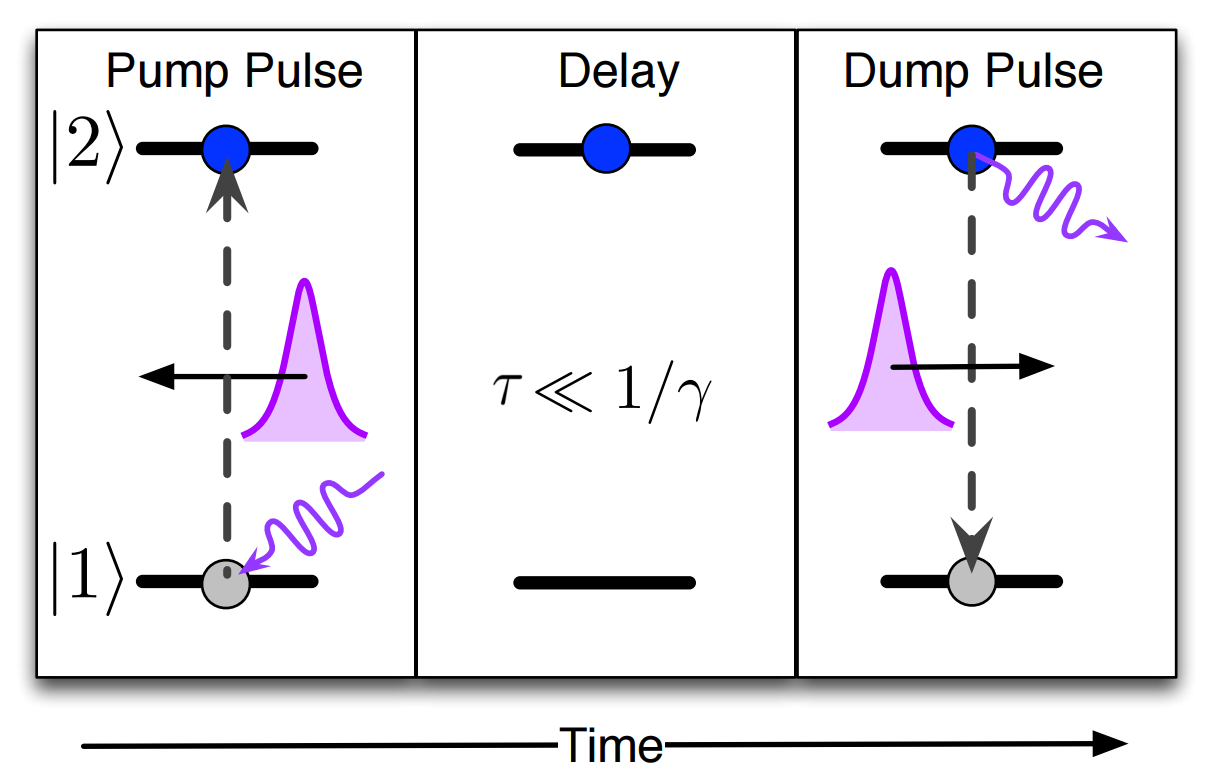}
    \caption{Pulse sequence of the stimulated force. As described below, the delay time $\tau$ between pump and dump pulses is chosen to be much smaller than the spontaneous emission lifetime ($\tau\ll1/\gamma$).}
    \label{fig:pulse_sequence}
\end{figure}

In this Letter, we demonstrate and benchmark an optical force derived entirely from stimulated scattering of mode-locked (ML) laser pulses \cite{Kazantsev1974TheAcceleration,Jayich2014theoryProposal}, shown in Fig.~\ref{fig:pulse_sequence}. Early work on this technique showed order-of-unity force gains over spontaneous scattering \cite{Voitsekhovich1994Observation,Nolle1996AtomicBeam,Goepfert1997Stimulated}. Here, by using a pre-cooled sample of atoms to benchmark and optimize the force, we show that its performance can be substantially improved. We are able to achieve an average of $\Upsilon = (19^{+6}_{-4})$ momentum transfers of $\hbar k$ per spontaneous emission event. This potentially extends optical deceleration to molecules with state leakage probabilities an order of magnitude worse than currently used species, such as complex polyatomics \cite{Kozyryev2016Proposal} and molecules well-suited to precision measurement \cite{Kozyryev2017PrecisionMeasurement,Denis2019Enhancement,Lim2018YbF,Hunter2012TlF}.

The stimulated force we demonstrate here is generated by the fast repetition of a cycle in which a time-ordered, counterpropagating pair of picosecond laser pulses (``$\pi$-pulses'') illuminate the sample. As illustrated in Fig.~\ref{fig:pulse_sequence} (see also \cite{Kazantsev1974TheAcceleration,Jayich2014theoryProposal,Voitsekhovich1994Observation,Nolle1996AtomicBeam,Goepfert1997Stimulated}), a ground-state molecule from a molecular beam is first excited by absorbing a photon from the ``pump pulse" that is counter-propagating with respect to the molecular beam, thereby losing momentum $\hbar k$. The molecule is then immediately illuminated by a co-propagating ``dump pulse," which deterministically drives the molecule back to its original ground state via stimulated emission and removes another $\hbar k$ of momentum from the molecule. The direction of the force is set by the order in which the pulses arrive, which introduces the necessary asymmetry to establish a preferred direction.  This cycle can be repeated rapidly to create an approximately continuous deceleration force that can be much stronger than spontaneous scattering.  The broad spectrum coverage of the ultrafast laser pulses allows for simultaneous deceleration of molecules from a wide range of velocities, and further augmentation of this scheme with adiabatic rapid passage and single-photon cooling has been studied theoretically \cite{Jayich2014theoryProposal}.

\begin{figure}
    \centering
    \includegraphics[scale=1.2]{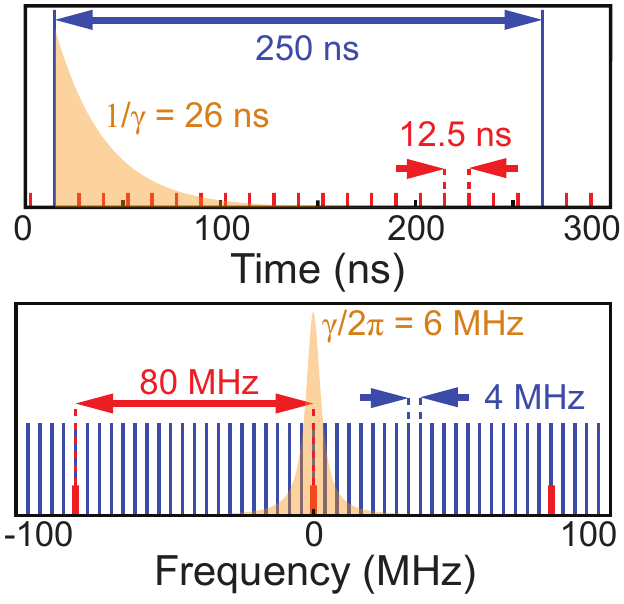}
    \caption{Time domain (upper) and frequency domain (lower) illustration of single-beam processes in this work. The ML laser generates $30 \mbox{ ps}$ pulses at $12.5 \mbox{ ns}$ intervals ($80 \mbox{ MHz}$, red). A Pockels cell increases this inter-pulse delay to $250 \mbox{ ns}$ ($4 \mbox{ MHz}$, blue) to ensure $>99.99\%$ decay probability between pulses. The excited state probability for an atom excited by the first pulse is represented by the yellow area in the time domain figure. The corresponding atomic spectrum is shown in the frequency domain figure.}
    \label{fig:domain}
\end{figure}

We demonstrate and benchmark this force on a MOT of $10^7$ pre-cooled $(120\pm10)\mbox{ }\mu\mbox{K}$ ${}^{85}$Rb atoms using the ${}^2S_{1/2} \! \rightarrow \! {}^2P_{3/2}$ transition.  As illustrated in Fig.~\ref{fig:domain}, the ML laser pulses are generated from a Ti:sapphire laser emitting $30\mbox{ ps}$ pulses at 780~nm at a repetition period of $12.5\mbox{ ns}$ (equivalently a repetition frequency of 80 MHz).  Since we work with a cold sample that has effectively no Doppler broadening, we find that inter-pulse-pair coherence can lead to velocity-selective frequency comb tooth effects that will not be available for decelerating a Doppler-broadened sample such as a molecular beam. To keep these continuous-wave (cw) -like systematic effects from contributing to our measured force ($\gamma/2 = 19\times 10^6\mbox{ s}^{-1}$, where the comb tooth visibility scales as $V \!\propto \mathrm{sech}\!\left( T_\mathrm{rep} \gamma/2\right)$ \cite{Ip2018Phonon}), a Pockels cell is used for pulse picking (power extinction ratio $= 7\times 10^{-3}$)
, increasing the pulse-to-pulse separation time from $12.5\mbox{ ns}$ to $T_\mathrm{rep}= 250\mbox{ ns}$. The $1/\mathrm{e}^2$ intensity diameter of the ML laser beams at the position of atomic cloud is $w=(0.65\pm0.03)$~mm, but non-gaussian variations are also present, as discussed below.  

 The MOT light and magnetic fields are turned off before the ML laser pulse trains are introduced. The dump beam path is made $\approx\!10\mbox{ cm}$ longer than the pump beam path to set a $\tau =(310\pm60)\mbox{ ps}$ intra-pulse-pair delay. This delay distance ensures no temporal overlap between pump and dump pulses, and can be reduced to nearly the pulse duration if it becomes a limiting factor in applications.

\begin{figure}
\includegraphics[width=0.95\columnwidth]{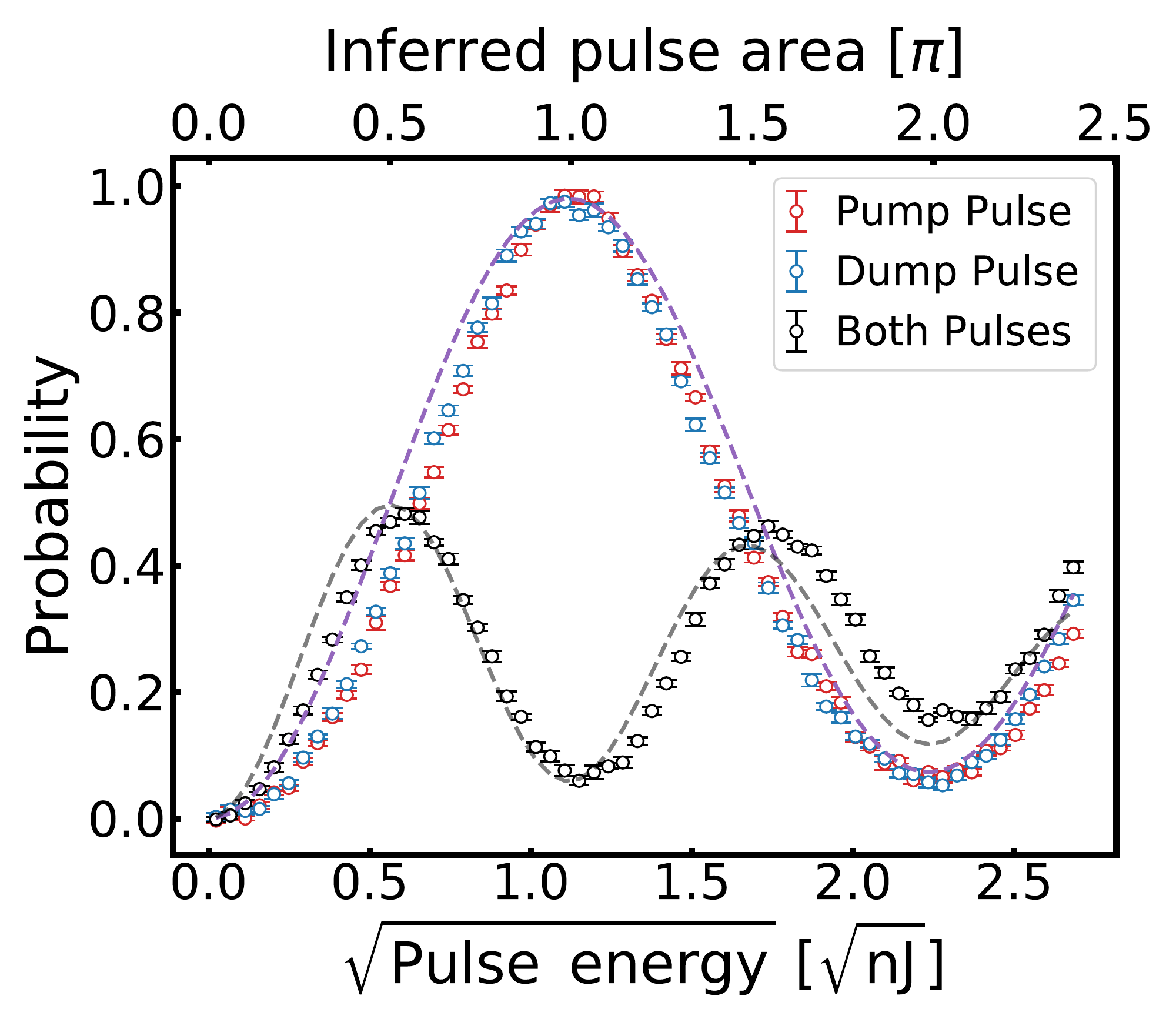}
\caption{Coherent Rabi flops on an optical frequency electric dipole transition.  The oscillation period allows identification and matching of the $\pi$-pulse pulse energy for both beams.  Fluorescence is collected from the atom cloud transverse to the acceleration direction for single pulses from the pump beam (red), dump beam (blue), or a pump-then-dump sequence (black).  The $\pi$-pulse condition is satisfied slightly above $1\mbox{ nJ}$, and the probability of spontaneous emission (the vertical axis) is calibrated from the measurements using Eqs.~\ref{eq:Pex1} and \ref{eq:Ngamma}, which are shown as dashed curves.}
\label{fig:rabi_flop}
\end{figure}

To calibrate and match the effective average pulse fluence from both beams, the atoms are illuminated with single pulses as the pulse energy is scanned, shown in Fig.~\ref{fig:rabi_flop}.  Fluorescence from spontaneous decays collected perpendicular to the ML beam propagation direction shows clear, coherent Rabi flops, an observation that is made possible despite the short lifetime of this transition ($1/\gamma\! = \!26\mbox{ ns}$) by the ultrafast excitation. The period of the Rabi flops allows us to infer the pulse area, shown as the top axis in Fig.~\ref{fig:rabi_flop}. We model the excited state probability ($P(\theta_\mathrm{o})$) as a function of pulse area $\theta \equiv \int\! \mathrm{d}t \,\Omega(t) \propto \sqrt{\mbox{pulse energy}}$ as coherent evolution averaged over a normal distribution of pulse areas with average value $\theta_\mathrm{o}$ and standard deviation $\sigma_\theta$,
\begin{equation}
P(\theta_\mathrm{o}) = \frac{1}{2} \left(1 - \mathrm{e}^{-\sigma_\theta^2/2} \cos \left( \theta_\mathrm{o} \right)  \right), \label{eq:Pex1}
\end{equation}
shown as a dashed, purple curve in Fig.~\ref{fig:rabi_flop}.

When the sequential, counter-propagating pump-then-dump pulses illuminate the atoms, the relative phase between them is spatially dependent on a length scale of a fraction of an optical wavelength.  Since the atom cloud is large compared to $\lambda$, we model the ensemble-averaged interaction as devoid of intra-pulse-pair coherence.  The excited state probability after the pump-dump sequence is given by
\begin{eqnarray}
P_\mathrm{ex}^{(2)}(\theta_\mathrm{o}) &=& \frac{1}{2}\left(1 - \mathrm{e}^{-\sigma_\theta^2/2} \cos \left( \theta_\mathrm{o} \right) \right) -\frac{1}{2} \mathrm{e}^{-\gamma\tau} \times \nonumber \\
&& \!\!\!\!\!\!\!\!\!\!\!\!\!\!\!\! \left(\frac{1}{2} - \mathrm{e}^{-\sigma_\theta^2/2} \cos \left( \theta_\mathrm{o} \right) + \frac{1}{2} \mathrm{e}^{-2\sigma_\theta^2} \cos \left(2 \theta_\mathrm{o} \right)  \right),\label{eq:Pex2}
\end{eqnarray}
where we assume that the two pulses have the same pulse area and $\tau$ is the time delay between the pulses.  Eq.~\ref{eq:Pex2} is combined with the probability of spontaneous emission between the pump and the dump pulse to give the expected number of spontaneous emissions per pulse pair,
\begin{equation}
    \left\langle N_{\gamma}\right\rangle = 1 - \frac{3}{4}\rme^{-\gamma \tau} + (1 - \rme^{-\gamma \tau})(2 \bar{P} - 1) - \frac{1}{4} \rme^{-\gamma \tau} (2 \bar{P} - 1)^4,\label{eq:Ngamma}
\end{equation}
where $\bar{P}\equiv P(\theta_{\mathrm{o}}\!=\!\pi)$ is the average single-pulse population transfer fidelity at the $\pi$-pulse condition from Eq.~\ref{eq:Pex1}.  Here we assume that any residual excited state population decays before the next pulse pair, the probability of which was made greater than $0.9999$ by the pulse picking to avoid frequency comb effects.  

The $\pi$-pulse condition is determined by finding the maximum of the single pulse (average fluorescence of pump-only and dump-only pulses) and local minimum of the pump-then-dump fluorescence signals, respectively, which coincide at the same pulse area. The measured $\pi$-pulse energy agrees with the theoretical prediction for a transform-limited, $30\mbox{ ps}$ pulse to within 20\%, confirming that the laser pulses in this work are nearly transform limited. Using Eq.~\ref{eq:Pex1} and Eq.~\ref{eq:Ngamma}, the ratio between the detected single-pulse and pump-then-dump fluorescence signals at the $\pi$-pulse condition does not require a calibration of the fluorescence collection efficiency, and provides a stand-alone measurement of $\sigma_\theta/\theta_\mathrm{o} = (0.09 \pm 0.01)$. This measured relative standard deviation is significantly higher than the relative pulse area variance measured from shot-to-shot pulse energy fluctuations ($\approx0.01$), suggesting that it is due to the transverse intensity variation of the beam across the atomic sample. Since this intensity gradient is static, we expect the average pulse fluence experienced by an atom to depend systematically on its position, and certain positions should experience repeatable population transfer fidelities that are significantly better than the average over the cloud.

The necessary calibration factor between fluorescence signal and population transfer fidelity is provided from Eq.~\ref{eq:Pex1} once $\sigma_\theta$ is known, and Fig.~\ref{fig:rabi_flop} shows that the average $\pi$-pulse population transfer fidelity is $\approx 98\%$ for each beam. As shown in Fig.~\ref{fig:rabi_flop}, this model overshoots the data for intermediate pulse areas ($\theta_\mathrm{o} \neq n \pi$), which we believe is caused by the finite optical depth of the sample leading to preferential emission in the forward direction \cite{Verne}.  Since the actual force is implemented at the $\pi$-pulse condition, we are primarily concerned with the agreement between our model and the data at this location. As a way to check the consistency of this model, we compare the predicted vs.~measured fluorescence at the $2\pi$-pulse condition for a single beam; the measured excited state fraction is $(5.9\pm0.8)\%$, in agreement with our model prediction of $(7\pm2)\%$ from Eq.~\ref{eq:Ngamma}.

Using this model, the efficiency of momentum transfer per pulse pair is 
\begin{equation}
\frac{\langle \Delta p \rangle}{2\hbar k} =  \frac{1}{8}\rme^{-\gamma \tau} \left( 3 + 4(2\bar{P} - 1) + (2 \bar{P}-1)^4 \right). \label{eq:Deltap}
\end{equation}
The assumption of full spontaneous emission of any excited population before the next pulse pair yields an expression for the number of $\hbar k$ photon momenta (along the pump beam propagation direction) transferred per spontaneous emission event,
\begin{equation}
\Upsilon = \left( \frac{8 \bar{P}}{\rme^{-\gamma \tau} \left( 3 + 4(2\bar{P}-1) + (2\bar{P}-1)^4\right)} - 1\right)^{-1}.\label{eq:Upsilon}
\end{equation}
Taking $\bar{P}=(0.980 \pm 0.005)$ from the data in Fig.~\ref{fig:rabi_flop} will yield the prediction $\Upsilon = (32 \pm 4)$. This prediction, however, does not account for potential intensity mismatch of the two beams in space or variations in $\bar{P}$ that may appear as atoms are accelerated along the beam. Since Eq.~\ref{eq:Upsilon} diverges as $\bar{P}\!\!\rightarrow\!\!1$ (for $\tau =0 $), small additional imperfections in pulse area may significantly decrease $\Upsilon$.

\begin{figure}
	\centering
        \includegraphics[width=0.95\columnwidth]{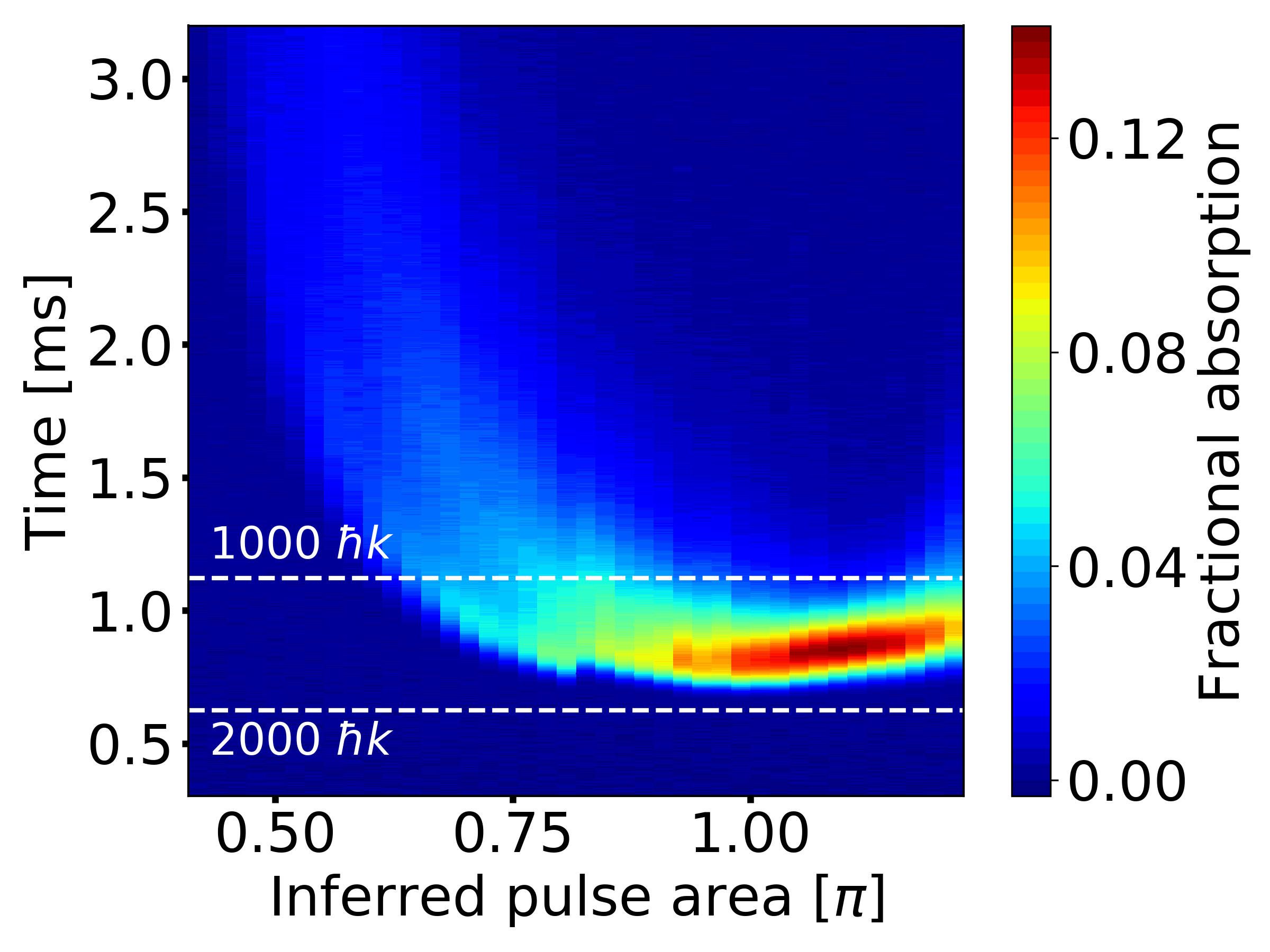}
    \caption{Effect of varying pulse energy on the arrival times of the atoms at the TOF detection position. Each vertical cross-section is a TOF trace (see \cite{Supplementary}). The dashed guide lines represent the theoretical arrival times if the indicated momentum had been transfered to the atoms.  These diagnostic data were taken before optimizing the force, and the arrival times of the fastest $10\%$ atoms corresponds to $\Upsilon \approx 6$.}
     \label{fig:meshplot}
\end{figure}

To obtain a better measurement of $\Upsilon$, we benchmark the momentum transfer itself by applying 1000 repeated pump-dump sequences (i.e.~2000 total pulses) to the atoms followed by TOF measurements that allow us to determine the time-averaged force.  A resonant, cw laser beam centered $4\!-\!6\mbox{ mm}$ away from the initial position of the atomic cloud in the direction of the stimulated force is used to record absorption as a function of time for atoms accelerated by the pump-dump pulse pairs. Fig.~\ref{fig:meshplot} shows TOF measurements at different pulse energies. In proximity to the $\pi$-pulse energy, the stimulated force becomes more efficient, resulting in better acceleration and earlier arrival times. Further, as the population transfer fidelity nears 1, the arrival time distribution of the atoms narrows, as expected from the reduction in quantum projection noise associated with the outcome of each pulse pair becoming more deterministic.  Even so, the width of the distribution at the $\pi$-pulse condition far exceeds what would be expected for uncorrelated pulse area fluctuations (which would contribute an arrival time spread of order $10\mbox{ }\mu\mbox{s}$, compared to the observed width of order $100\mbox{ }\mu\mbox{s}$), and is instead caused by the systematic variations in population transfer fidelity associated with the nonuniform transverse intensity profiles of the beams.

To quantify the acceleration, TOF measurements are used to optimize the force and are then performed at 5 different absorption beam locations along the trajectory of the accelerated atoms to control for the initial position and its spread (see \cite{Supplementary}). As discussed above, since the spatial profile of the pulses is not perfectly uniform, some locations in space experience \textit{systematically} higher population transfer fidelity than others, and it is these that represent the ensemble most interesting for considering future applications to molecules.  During optimization of the beam positions and strengths, the arrival time of the fastest moving $10\%$ of the atoms was minimized.  For these atoms, we obtain a velocity of $(11.0 \pm 0.3)\mbox{ m}/\mbox{s}$, corresponding to a total momentum transfer of $(1820\pm50) \hbar k$ from 2000 pulses and a momentum transfer efficiency of $(91\pm3)\%$.  Using this momentum transfer efficiency, Eq.~\ref{eq:Deltap} can be solved for the effective average $\pi$-pulse population transfer fidelity, yielding $\bar{P} = (0.958 \pm 0.014)$, which gives the measured figure of merit $\Upsilon = 19\;{}^{+6}_{-4}$. The lower values of $\bar{P}$ and $\Upsilon$ measured from \textit{in situ} acceleration measurements as compared to those inferred from few-pulse fluorescence experiments (e.g. Fig.~\ref{fig:rabi_flop}) highlight the need to perform measurements of this kind by measuring the actual momentum transfer, which is sensitive to more potential systematic effects than observations of internal state dynamics. For instance, due to the finite extinction ratio of the pulse picker, small comb tooth effects can become important between pulse pairs during long pushing sequences.  We find that the momentum transfer efficiency is insensitive to the comb tooth position as long as the nearest-resonant tooth is not close to the the atomic transition, a condition that is maintained for these experiments with the passive stability of the laser (see \cite{Supplementary}).

Comparison of this measurement of $\Upsilon$ to other methods in the literature is complicated by the fact that very few demonstrations of stimulated slowing techniques report the average gains in cycle closure that they are designed to provide (though a recent demonstration of the bichromatic force on polyatomic molecules \cite{Kozyryev2018BichromaticForce} achieved $\Upsilon=(3.7\pm0.7)$ \cite{Kozyryev2018PrivateComminucation}). Two other performance indicators are more common: the excited state fraction, which determines the ensemble-averaged radiative decay rate, and the force gain factor, which is the ratio of the magnitude of the stimulated force over the theoretical maximum radiative force for an ideal two-level system. The time-averaged excited state fraction induced by the bichromatic force for a two-level system can be optimized to $41\%$, though it could be improved further to $24\%$ with a four-color force scheme \cite{Galica2013four_color}.  The pulsed scheme in this work can be viewed as a polychromatic limit of the bichromatic force, and the time-averaged pump-then-dump excited state fraction achieved here is $(1.0\pm0.2)\%$. Likewise,  experimental work on bichromatic deflection has demonstrated a force gain factor improvement of 1.1 \cite{Kozyryev2018BichromaticForce} on polyatomic molecules and a similar value on diatomic molecules \cite{Galica2018BichromaticDeflection}, whereas spontaneous scattering force experiments on polyatomic \cite{Kozyryev2016PolyatomicRadiativeForce} and diatomic \cite{Shuman2009DiatomicRadiativeForce} molecules have shown force gain factors of 0.5 and 0.29 respectively. With the intentionally low repetition rate adopted in our measurement to eliminate comb tooth systematics, we nonetheless measure a force gain factor of $(0.38\pm0.01)$, already comparable to spontaneous scattering. 

One possible application of this scheme would be for laser deceleration of YbOH, a polyatomic molecule candidate for future measurements of the electron electric dipole moment \cite{Denis2019Enhancement}. White light slowing with five repump lasers has been proposed to produce a spontaneous scattering force sufficient for stopping a beam of YbOH \cite{Kozyryev2017PrecisionMeasurement}, whereas use of this pulsed stimulated optical force with $\Upsilon=19$ would reduce the number of repump lasers by three. In addition, assuming the demonstrated momentum transfer efficiency ($91\%$) can be achieved at $80 \mbox{ MHz}$, a YbOH beam can be slowed from a initial speed of $150 \mbox{ m}/\mbox{s}$ to a full stop in $22 \mbox{ cm}$, thereby suppressing molecular losses due to transverse motion and increasing the molecular flux. At higher repetition rates, comb tooth effects can potentially to appear, but these effects can be minimized by carefully maintaining the $\pi$-pulse condition.

For molecular slowing, the bandwidth of the ML pulses should be smaller than the rotational splitting, but can still exceed the scale of relevant Doppler shifts. For the example above, the rotational ground state splitting of YbOH is $14.7 \mbox{ GHz}$ \cite{Nakhate2018YbOHSpectrum}, but a bandwidth of 500 MHz is more than sufficient to cover the expected velocity range down to a full stop in the lab.

\begin{acknowledgments}
The authors acknowledge helpful discussions with Paul Hamilton, Eric Hudson, and Ivan Kozyryev. This work was supported by the NSF CAREER Program under award No. 1455357.
\end{acknowledgments}

\bibliography{bibliography}

\begin{thebibliography}{63}%
\makeatletter
\providecommand \@ifxundefined [1]{%
 \@ifx{#1\undefined}
}%
\providecommand \@ifnum [1]{%
 \ifnum #1\expandafter \@firstoftwo
 \else \expandafter \@secondoftwo
 \fi
}%
\providecommand \@ifx [1]{%
 \ifx #1\expandafter \@firstoftwo
 \else \expandafter \@secondoftwo
 \fi
}%
\providecommand \natexlab [1]{#1}%
\providecommand \enquote  [1]{``#1''}%
\providecommand \bibnamefont  [1]{#1}%
\providecommand \bibfnamefont [1]{#1}%
\providecommand \citenamefont [1]{#1}%
\providecommand \href@noop [0]{\@secondoftwo}%
\providecommand \href [0]{\begingroup \@sanitize@url \@href}%
\providecommand \@href[1]{\@@startlink{#1}\@@href}%
\providecommand \@@href[1]{\endgroup#1\@@endlink}%
\providecommand \@sanitize@url [0]{\catcode `\\12\catcode `\$12\catcode
  `\&12\catcode `\#12\catcode `\^12\catcode `\_12\catcode `\%12\relax}%
\providecommand \@@startlink[1]{}%
\providecommand \@@endlink[0]{}%
\providecommand \url  [0]{\begingroup\@sanitize@url \@url }%
\providecommand \@url [1]{\endgroup\@href {#1}{\urlprefix }}%
\providecommand \urlprefix  [0]{URL }%
\providecommand \Eprint [0]{\href }%
\providecommand \doibase [0]{http://dx.doi.org/}%
\providecommand \selectlanguage [0]{\@gobble}%
\providecommand \bibinfo  [0]{\@secondoftwo}%
\providecommand \bibfield  [0]{\@secondoftwo}%
\providecommand \translation [1]{[#1]}%
\providecommand \BibitemOpen [0]{}%
\providecommand \bibitemStop [0]{}%
\providecommand \bibitemNoStop [0]{.\EOS\space}%
\providecommand \EOS [0]{\spacefactor3000\relax}%
\providecommand \BibitemShut  [1]{\csname bibitem#1\endcsname}%
\let\auto@bib@innerbib\@empty
\bibitem [{\citenamefont {Anderson}\ \emph {et~al.}(1995)\citenamefont
  {Anderson}, \citenamefont {Ensher}, \citenamefont {Matthews}, \citenamefont
  {Wieman},\ and\ \citenamefont {Cornell}}]{Anderson198BoseEinsteinCondensate}%
  \BibitemOpen
  \bibfield  {author} {\bibinfo {author} {\bibfnamefont {M.~H.}\ \bibnamefont
  {Anderson}}, \bibinfo {author} {\bibfnamefont {J.~R.}\ \bibnamefont
  {Ensher}}, \bibinfo {author} {\bibfnamefont {M.~R.}\ \bibnamefont
  {Matthews}}, \bibinfo {author} {\bibfnamefont {C.~E.}\ \bibnamefont
  {Wieman}}, \ and\ \bibinfo {author} {\bibfnamefont {E.~A.}\ \bibnamefont
  {Cornell}},\ }\href {\doibase 10.1126/science.269.5221.198} {\bibfield
  {journal} {\bibinfo  {journal} {Science}\ }\textbf {\bibinfo {volume}
  {269}},\ \bibinfo {pages} {198} (\bibinfo {year} {1995})}\BibitemShut
  {NoStop}%
\bibitem [{\citenamefont {Ludlow}\ \emph {et~al.}(2015)\citenamefont {Ludlow},
  \citenamefont {Boyd}, \citenamefont {Ye}, \citenamefont {Peik},\ and\
  \citenamefont {Schmidt}}]{Ludlow2015Clock}%
  \BibitemOpen
  \bibfield  {author} {\bibinfo {author} {\bibfnamefont {A.~D.}\ \bibnamefont
  {Ludlow}}, \bibinfo {author} {\bibfnamefont {M.~M.}\ \bibnamefont {Boyd}},
  \bibinfo {author} {\bibfnamefont {J.}~\bibnamefont {Ye}}, \bibinfo {author}
  {\bibfnamefont {E.}~\bibnamefont {Peik}}, \ and\ \bibinfo {author}
  {\bibfnamefont {P.~O.}\ \bibnamefont {Schmidt}},\ }\href {\doibase
  10.1103/RevModPhys.87.637} {\bibfield  {journal} {\bibinfo  {journal} {Rev.
  Mod. Phys.}\ }\textbf {\bibinfo {volume} {87}},\ \bibinfo {pages} {637}
  (\bibinfo {year} {2015})}\BibitemShut {NoStop}%
\bibitem [{\citenamefont {Cronin}\ \emph {et~al.}(2009)\citenamefont {Cronin},
  \citenamefont {Schmiedmayer},\ and\ \citenamefont
  {Pritchard}}]{Cronin2009interferometer}%
  \BibitemOpen
  \bibfield  {author} {\bibinfo {author} {\bibfnamefont {A.~D.}\ \bibnamefont
  {Cronin}}, \bibinfo {author} {\bibfnamefont {J.}~\bibnamefont
  {Schmiedmayer}}, \ and\ \bibinfo {author} {\bibfnamefont {D.~E.}\
  \bibnamefont {Pritchard}},\ }\href {\doibase 10.1103/RevModPhys.81.1051}
  {\bibfield  {journal} {\bibinfo  {journal} {Rev. Mod. Phys.}\ }\textbf
  {\bibinfo {volume} {81}},\ \bibinfo {pages} {1051} (\bibinfo {year}
  {2009})}\BibitemShut {NoStop}%
\bibitem [{\citenamefont {H{\"a}ffner}\ \emph {et~al.}(2008)\citenamefont
  {H{\"a}ffner}, \citenamefont {Roos},\ and\ \citenamefont
  {Blatt}}]{Haffner2008Quantum}%
  \BibitemOpen
  \bibfield  {author} {\bibinfo {author} {\bibfnamefont {H.}~\bibnamefont
  {H{\"a}ffner}}, \bibinfo {author} {\bibfnamefont {C.}~\bibnamefont {Roos}}, \
  and\ \bibinfo {author} {\bibfnamefont {R.}~\bibnamefont {Blatt}},\ }\href
  {\doibase https://doi.org/10.1016/j.physrep.2008.09.003} {\bibfield
  {journal} {\bibinfo  {journal} {Physics Reports}\ }\textbf {\bibinfo {volume}
  {469}},\ \bibinfo {pages} {155 } (\bibinfo {year} {2008})}\BibitemShut
  {NoStop}%
\bibitem [{\citenamefont {Wilkening}\ \emph {et~al.}(1984)\citenamefont
  {Wilkening}, \citenamefont {Ramsey},\ and\ \citenamefont
  {Larson}}]{Wilkening1984PTviolation}%
  \BibitemOpen
  \bibfield  {author} {\bibinfo {author} {\bibfnamefont {D.~A.}\ \bibnamefont
  {Wilkening}}, \bibinfo {author} {\bibfnamefont {N.~F.}\ \bibnamefont
  {Ramsey}}, \ and\ \bibinfo {author} {\bibfnamefont {D.~J.}\ \bibnamefont
  {Larson}},\ }\href {\doibase 10.1103/PhysRevA.29.425} {\bibfield  {journal}
  {\bibinfo  {journal} {Phys. Rev. A}\ }\textbf {\bibinfo {volume} {29}},\
  \bibinfo {pages} {425} (\bibinfo {year} {1984})}\BibitemShut {NoStop}%
\bibitem [{\citenamefont {Hinds}(1997)}]{Hinds1997testingTimeReversalSymmetry}%
  \BibitemOpen
  \bibfield  {author} {\bibinfo {author} {\bibfnamefont {E.~A.}\ \bibnamefont
  {Hinds}},\ }\href {http://stacks.iop.org/1402-4896/1997/i=T70/a=005}
  {\bibfield  {journal} {\bibinfo  {journal} {Physica Scripta}\ }\textbf
  {\bibinfo {volume} {1997}},\ \bibinfo {pages} {34} (\bibinfo {year}
  {1997})}\BibitemShut {NoStop}%
\bibitem [{\citenamefont {Vutha}\ \emph {et~al.}(2010)\citenamefont {Vutha},
  \citenamefont {Campbell}, \citenamefont {Gurevich}, \citenamefont {Hutzler},
  \citenamefont {Parsons}, \citenamefont {Patterson}, \citenamefont {Petrik},
  \citenamefont {Spaun}, \citenamefont {Doyle}, \citenamefont {Gabrielse},\
  and\ \citenamefont {DeMille}}]{Vutha2010EDMTiO}%
  \BibitemOpen
  \bibfield  {author} {\bibinfo {author} {\bibfnamefont {A.~C.}\ \bibnamefont
  {Vutha}}, \bibinfo {author} {\bibfnamefont {W.~C.}\ \bibnamefont {Campbell}},
  \bibinfo {author} {\bibfnamefont {Y.~V.}\ \bibnamefont {Gurevich}}, \bibinfo
  {author} {\bibfnamefont {N.~R.}\ \bibnamefont {Hutzler}}, \bibinfo {author}
  {\bibfnamefont {M.}~\bibnamefont {Parsons}}, \bibinfo {author} {\bibfnamefont
  {D.}~\bibnamefont {Patterson}}, \bibinfo {author} {\bibfnamefont
  {E.}~\bibnamefont {Petrik}}, \bibinfo {author} {\bibfnamefont
  {B.}~\bibnamefont {Spaun}}, \bibinfo {author} {\bibfnamefont {J.~M.}\
  \bibnamefont {Doyle}}, \bibinfo {author} {\bibfnamefont {G.}~\bibnamefont
  {Gabrielse}}, \ and\ \bibinfo {author} {\bibfnamefont {D.}~\bibnamefont
  {DeMille}},\ }\href {http://stacks.iop.org/0953-4075/43/i=7/a=074007}
  {\bibfield  {journal} {\bibinfo  {journal} {Journal of Physics B: Atomic,
  Molecular and Optical Physics}\ }\textbf {\bibinfo {volume} {43}},\ \bibinfo
  {pages} {074007} (\bibinfo {year} {2010})}\BibitemShut {NoStop}%
\bibitem [{\citenamefont {Altunta\ifmmode~\mbox{\c{s}}\else \c{s}\fi{}}\ \emph
  {et~al.}(2018)\citenamefont {Altunta\ifmmode~\mbox{\c{s}}\else \c{s}\fi{}},
  \citenamefont {Ammon}, \citenamefont {Cahn},\ and\ \citenamefont
  {DeMille}}]{Altunta2018NuclearSpinParityViolation}%
  \BibitemOpen
  \bibfield  {author} {\bibinfo {author} {\bibfnamefont {E.}~\bibnamefont
  {Altunta\ifmmode~\mbox{\c{s}}\else \c{s}\fi{}}}, \bibinfo {author}
  {\bibfnamefont {J.}~\bibnamefont {Ammon}}, \bibinfo {author} {\bibfnamefont
  {S.~B.}\ \bibnamefont {Cahn}}, \ and\ \bibinfo {author} {\bibfnamefont
  {D.}~\bibnamefont {DeMille}},\ }\href {\doibase
  10.1103/PhysRevLett.120.142501} {\bibfield  {journal} {\bibinfo  {journal}
  {Phys. Rev. Lett.}\ }\textbf {\bibinfo {volume} {120}},\ \bibinfo {pages}
  {142501} (\bibinfo {year} {2018})}\BibitemShut {NoStop}%
\bibitem [{\citenamefont {Kozyryev}\ and\ \citenamefont
  {Hutzler}(2017)}]{Kozyryev2017PrecisionMeasurement}%
  \BibitemOpen
  \bibfield  {author} {\bibinfo {author} {\bibfnamefont {I.}~\bibnamefont
  {Kozyryev}}\ and\ \bibinfo {author} {\bibfnamefont {N.~R.}\ \bibnamefont
  {Hutzler}},\ }\href {\doibase 10.1103/PhysRevLett.119.133002} {\bibfield
  {journal} {\bibinfo  {journal} {Phys. Rev. Lett.}\ }\textbf {\bibinfo
  {volume} {119}},\ \bibinfo {pages} {133002} (\bibinfo {year}
  {2017})}\BibitemShut {NoStop}%
\bibitem [{\citenamefont {DeMille}(2002)}]{Demille2002QuantumComputing}%
  \BibitemOpen
  \bibfield  {author} {\bibinfo {author} {\bibfnamefont {D.}~\bibnamefont
  {DeMille}},\ }\href {\doibase 10.1103/PhysRevLett.88.067901} {\bibfield
  {journal} {\bibinfo  {journal} {Phys. Rev. Lett.}\ }\textbf {\bibinfo
  {volume} {88}},\ \bibinfo {pages} {067901} (\bibinfo {year}
  {2002})}\BibitemShut {NoStop}%
\bibitem [{\citenamefont {Andr{\'e}}\ \emph {et~al.}(2006)\citenamefont
  {Andr{\'e}}, \citenamefont {DeMille}, \citenamefont {Doyle}, \citenamefont
  {Lukin}, \citenamefont {Maxwell}, \citenamefont {Rabl}, \citenamefont
  {Schoelkopf},\ and\ \citenamefont {Zoller}}]{Andre2006QIinterface}%
  \BibitemOpen
  \bibfield  {author} {\bibinfo {author} {\bibfnamefont {A.}~\bibnamefont
  {Andr{\'e}}}, \bibinfo {author} {\bibfnamefont {D.}~\bibnamefont {DeMille}},
  \bibinfo {author} {\bibfnamefont {J.~M.}\ \bibnamefont {Doyle}}, \bibinfo
  {author} {\bibfnamefont {M.~D.}\ \bibnamefont {Lukin}}, \bibinfo {author}
  {\bibfnamefont {S.~E.}\ \bibnamefont {Maxwell}}, \bibinfo {author}
  {\bibfnamefont {P.}~\bibnamefont {Rabl}}, \bibinfo {author} {\bibfnamefont
  {R.~J.}\ \bibnamefont {Schoelkopf}}, \ and\ \bibinfo {author} {\bibfnamefont
  {P.}~\bibnamefont {Zoller}},\ }\href {http://dx.doi.org/10.1038/nphys386}
  {\bibfield  {journal} {\bibinfo  {journal} {Nature Physics}\ }\textbf
  {\bibinfo {volume} {2}},\ \bibinfo {pages} {636 EP } (\bibinfo {year}
  {2006})},\ \bibinfo {note} {article}\BibitemShut {NoStop}%
\bibitem [{\citenamefont {Rabl}\ \emph {et~al.}(2006)\citenamefont {Rabl},
  \citenamefont {DeMille}, \citenamefont {Doyle}, \citenamefont {Lukin},
  \citenamefont {Schoelkopf},\ and\ \citenamefont
  {Zoller}}]{Rabl2006QuantumMemory}%
  \BibitemOpen
  \bibfield  {author} {\bibinfo {author} {\bibfnamefont {P.}~\bibnamefont
  {Rabl}}, \bibinfo {author} {\bibfnamefont {D.}~\bibnamefont {DeMille}},
  \bibinfo {author} {\bibfnamefont {J.~M.}\ \bibnamefont {Doyle}}, \bibinfo
  {author} {\bibfnamefont {M.~D.}\ \bibnamefont {Lukin}}, \bibinfo {author}
  {\bibfnamefont {R.~J.}\ \bibnamefont {Schoelkopf}}, \ and\ \bibinfo {author}
  {\bibfnamefont {P.}~\bibnamefont {Zoller}},\ }\href {\doibase
  10.1103/PhysRevLett.97.033003} {\bibfield  {journal} {\bibinfo  {journal}
  {Phys. Rev. Lett.}\ }\textbf {\bibinfo {volume} {97}},\ \bibinfo {pages}
  {033003} (\bibinfo {year} {2006})}\BibitemShut {NoStop}%
\bibitem [{\citenamefont {Hudson}\ and\ \citenamefont
  {Campbell}(2018)}]{Hudson2018QuantumLogic}%
  \BibitemOpen
  \bibfield  {author} {\bibinfo {author} {\bibfnamefont {E.~R.}\ \bibnamefont
  {Hudson}}\ and\ \bibinfo {author} {\bibfnamefont {W.~C.}\ \bibnamefont
  {Campbell}},\ }\href@noop {} {\bibfield  {journal} {\bibinfo  {journal}
  {Phys. Rev. A}\ }\textbf {\bibinfo {volume} {98}},\ \bibinfo {pages}
  {040302(R)} (\bibinfo {year} {2018})}\BibitemShut {NoStop}%
\bibitem [{\citenamefont {Baranov}\ \emph {et~al.}(2012)\citenamefont
  {Baranov}, \citenamefont {Dalmonte}, \citenamefont {Pupillo},\ and\
  \citenamefont {Zoller}}]{Baranov2012QuantumSimulation}%
  \BibitemOpen
  \bibfield  {author} {\bibinfo {author} {\bibfnamefont {M.~A.}\ \bibnamefont
  {Baranov}}, \bibinfo {author} {\bibfnamefont {M.}~\bibnamefont {Dalmonte}},
  \bibinfo {author} {\bibfnamefont {G.}~\bibnamefont {Pupillo}}, \ and\
  \bibinfo {author} {\bibfnamefont {P.}~\bibnamefont {Zoller}},\ }\href
  {\doibase 10.1021/cr2003568} {\bibfield  {journal} {\bibinfo  {journal}
  {Chemical Reviews}\ }\textbf {\bibinfo {volume} {112}},\ \bibinfo {pages}
  {5012} (\bibinfo {year} {2012})},\ \bibinfo {note} {pMID:
  22877362}\BibitemShut {NoStop}%
\bibitem [{\citenamefont {Micheli}\ \emph {et~al.}(2006)\citenamefont
  {Micheli}, \citenamefont {Brennen},\ and\ \citenamefont
  {Zoller}}]{Micheli2006QuantumSimulation}%
  \BibitemOpen
  \bibfield  {author} {\bibinfo {author} {\bibfnamefont {A.}~\bibnamefont
  {Micheli}}, \bibinfo {author} {\bibfnamefont {G.~K.}\ \bibnamefont
  {Brennen}}, \ and\ \bibinfo {author} {\bibfnamefont {P.}~\bibnamefont
  {Zoller}},\ }\href {http://dx.doi.org/10.1038/nphys287} {\bibfield  {journal}
  {\bibinfo  {journal} {Nature Physics}\ }\textbf {\bibinfo {volume} {2}},\
  \bibinfo {pages} {341 EP } (\bibinfo {year} {2006})},\ \bibinfo {note}
  {article}\BibitemShut {NoStop}%
\bibitem [{\citenamefont {Ospelkaus}\ \emph {et~al.}(2010)\citenamefont
  {Ospelkaus}, \citenamefont {Ni}, \citenamefont {Wang}, \citenamefont
  {de~Miranda}, \citenamefont {Neyenhuis}, \citenamefont {Qu{\'e}m{\'e}ner},
  \citenamefont {Julienne}, \citenamefont {Bohn}, \citenamefont {Jin},\ and\
  \citenamefont {Ye}}]{Ospelkaus2010QuantumChemistry}%
  \BibitemOpen
  \bibfield  {author} {\bibinfo {author} {\bibfnamefont {S.}~\bibnamefont
  {Ospelkaus}}, \bibinfo {author} {\bibfnamefont {K.-K.}\ \bibnamefont {Ni}},
  \bibinfo {author} {\bibfnamefont {D.}~\bibnamefont {Wang}}, \bibinfo {author}
  {\bibfnamefont {M.~H.~G.}\ \bibnamefont {de~Miranda}}, \bibinfo {author}
  {\bibfnamefont {B.}~\bibnamefont {Neyenhuis}}, \bibinfo {author}
  {\bibfnamefont {G.}~\bibnamefont {Qu{\'e}m{\'e}ner}}, \bibinfo {author}
  {\bibfnamefont {P.~S.}\ \bibnamefont {Julienne}}, \bibinfo {author}
  {\bibfnamefont {J.~L.}\ \bibnamefont {Bohn}}, \bibinfo {author}
  {\bibfnamefont {D.~S.}\ \bibnamefont {Jin}}, \ and\ \bibinfo {author}
  {\bibfnamefont {J.}~\bibnamefont {Ye}},\ }\href {\doibase
  10.1126/science.1184121} {\bibfield  {journal} {\bibinfo  {journal}
  {Science}\ }\textbf {\bibinfo {volume} {327}},\ \bibinfo {pages} {853}
  (\bibinfo {year} {2010})}\BibitemShut {NoStop}%
\bibitem [{\citenamefont {Krems}(2008)}]{Krems2008ColdControlledChemistry}%
  \BibitemOpen
  \bibfield  {author} {\bibinfo {author} {\bibfnamefont {R.~V.}\ \bibnamefont
  {Krems}},\ }\href {\doibase 10.1039/B802322K} {\bibfield  {journal} {\bibinfo
   {journal} {Phys. Chem. Chem. Phys.}\ }\textbf {\bibinfo {volume} {10}},\
  \bibinfo {pages} {4079} (\bibinfo {year} {2008})}\BibitemShut {NoStop}%
\bibitem [{\citenamefont {Barry}\ \emph {et~al.}(2014)\citenamefont {Barry},
  \citenamefont {McCarron}, \citenamefont {Norrgard}, \citenamefont
  {Steinecker},\ and\ \citenamefont
  {DeMille}}]{Barry2014MOTofDiatomicMolecule}%
  \BibitemOpen
  \bibfield  {author} {\bibinfo {author} {\bibfnamefont {J.~F.}\ \bibnamefont
  {Barry}}, \bibinfo {author} {\bibfnamefont {D.~J.}\ \bibnamefont {McCarron}},
  \bibinfo {author} {\bibfnamefont {E.~B.}\ \bibnamefont {Norrgard}}, \bibinfo
  {author} {\bibfnamefont {M.~H.}\ \bibnamefont {Steinecker}}, \ and\ \bibinfo
  {author} {\bibfnamefont {D.}~\bibnamefont {DeMille}},\ }\href
  {http://dx.doi.org/10.1038/nature13634} {\bibfield  {journal} {\bibinfo
  {journal} {Nature}\ }\textbf {\bibinfo {volume} {512}},\ \bibinfo {pages}
  {286 EP } (\bibinfo {year} {2014})}\BibitemShut {NoStop}%
\bibitem [{\citenamefont {Norrgard}\ \emph {et~al.}(2016)\citenamefont
  {Norrgard}, \citenamefont {McCarron}, \citenamefont {Steinecker},
  \citenamefont {Tarbutt},\ and\ \citenamefont
  {DeMille}}]{Norrgard2016RadioFrequencyMOT}%
  \BibitemOpen
  \bibfield  {author} {\bibinfo {author} {\bibfnamefont {E.~B.}\ \bibnamefont
  {Norrgard}}, \bibinfo {author} {\bibfnamefont {D.~J.}\ \bibnamefont
  {McCarron}}, \bibinfo {author} {\bibfnamefont {M.~H.}\ \bibnamefont
  {Steinecker}}, \bibinfo {author} {\bibfnamefont {M.~R.}\ \bibnamefont
  {Tarbutt}}, \ and\ \bibinfo {author} {\bibfnamefont {D.}~\bibnamefont
  {DeMille}},\ }\href {\doibase 10.1103/PhysRevLett.116.063004} {\bibfield
  {journal} {\bibinfo  {journal} {Phys. Rev. Lett.}\ }\textbf {\bibinfo
  {volume} {116}},\ \bibinfo {pages} {063004} (\bibinfo {year}
  {2016})}\BibitemShut {NoStop}%
\bibitem [{\citenamefont {Steinecker}\ \emph {et~al.}(2016)\citenamefont
  {Steinecker}, \citenamefont {McCarron}, \citenamefont {Zhu},\ and\
  \citenamefont {DeMille}}]{Steinecker2016ImprovedRadioFreqMOT}%
  \BibitemOpen
  \bibfield  {author} {\bibinfo {author} {\bibfnamefont {M.~H.}\ \bibnamefont
  {Steinecker}}, \bibinfo {author} {\bibfnamefont {D.~J.}\ \bibnamefont
  {McCarron}}, \bibinfo {author} {\bibfnamefont {Y.}~\bibnamefont {Zhu}}, \
  and\ \bibinfo {author} {\bibfnamefont {D.}~\bibnamefont {DeMille}},\ }\href
  {\doibase 10.1002/cphc.201600967} {\bibfield  {journal} {\bibinfo  {journal}
  {ChemPhysChem}\ }\textbf {\bibinfo {volume} {17}},\ \bibinfo {pages} {3664}
  (\bibinfo {year} {2016})}\BibitemShut {NoStop}%
\bibitem [{\citenamefont {Truppe}\ \emph
  {et~al.}(2017{\natexlab{a}})\citenamefont {Truppe}, \citenamefont {Williams},
  \citenamefont {Hambach}, \citenamefont {Caldwell}, \citenamefont {Fitch},
  \citenamefont {Hinds}, \citenamefont {Sauer},\ and\ \citenamefont
  {Tarbutt}}]{Truppe2017SubDopplerMOT}%
  \BibitemOpen
  \bibfield  {author} {\bibinfo {author} {\bibfnamefont {S.}~\bibnamefont
  {Truppe}}, \bibinfo {author} {\bibfnamefont {H.~J.}\ \bibnamefont
  {Williams}}, \bibinfo {author} {\bibfnamefont {M.}~\bibnamefont {Hambach}},
  \bibinfo {author} {\bibfnamefont {L.}~\bibnamefont {Caldwell}}, \bibinfo
  {author} {\bibfnamefont {N.~J.}\ \bibnamefont {Fitch}}, \bibinfo {author}
  {\bibfnamefont {E.~A.}\ \bibnamefont {Hinds}}, \bibinfo {author}
  {\bibfnamefont {B.~E.}\ \bibnamefont {Sauer}}, \ and\ \bibinfo {author}
  {\bibfnamefont {M.~R.}\ \bibnamefont {Tarbutt}},\ }\href
  {http://dx.doi.org/10.1038/nphys4241} {\bibfield  {journal} {\bibinfo
  {journal} {Nature Physics}\ }\textbf {\bibinfo {volume} {13}},\ \bibinfo
  {pages} {1173 EP } (\bibinfo {year} {2017}{\natexlab{a}})}\BibitemShut
  {NoStop}%
\bibitem [{\citenamefont {Anderegg}\ \emph {et~al.}(2017)\citenamefont
  {Anderegg}, \citenamefont {Augenbraun}, \citenamefont {Chae}, \citenamefont
  {Hemmerling}, \citenamefont {Hutzler}, \citenamefont {Ravi}, \citenamefont
  {Collopy}, \citenamefont {Ye}, \citenamefont {Ketterle},\ and\ \citenamefont
  {Doyle}}]{Anderegg2017HighDensityRadioFreqMOT}%
  \BibitemOpen
  \bibfield  {author} {\bibinfo {author} {\bibfnamefont {L.}~\bibnamefont
  {Anderegg}}, \bibinfo {author} {\bibfnamefont {B.~L.}\ \bibnamefont
  {Augenbraun}}, \bibinfo {author} {\bibfnamefont {E.}~\bibnamefont {Chae}},
  \bibinfo {author} {\bibfnamefont {B.}~\bibnamefont {Hemmerling}}, \bibinfo
  {author} {\bibfnamefont {N.~R.}\ \bibnamefont {Hutzler}}, \bibinfo {author}
  {\bibfnamefont {A.}~\bibnamefont {Ravi}}, \bibinfo {author} {\bibfnamefont
  {A.}~\bibnamefont {Collopy}}, \bibinfo {author} {\bibfnamefont
  {J.}~\bibnamefont {Ye}}, \bibinfo {author} {\bibfnamefont {W.}~\bibnamefont
  {Ketterle}}, \ and\ \bibinfo {author} {\bibfnamefont {J.~M.}\ \bibnamefont
  {Doyle}},\ }\href {\doibase 10.1103/PhysRevLett.119.103201} {\bibfield
  {journal} {\bibinfo  {journal} {Phys. Rev. Lett.}\ }\textbf {\bibinfo
  {volume} {119}},\ \bibinfo {pages} {103201} (\bibinfo {year}
  {2017})}\BibitemShut {NoStop}%
\bibitem [{\citenamefont {Anderegg}\ \emph {et~al.}(2018)\citenamefont
  {Anderegg}, \citenamefont {Augenbraun}, \citenamefont {Bao}, \citenamefont
  {Burchesky}, \citenamefont {Cheuk}, \citenamefont {Ketterle},\ and\
  \citenamefont {Doyle}}]{Anderegg2018LaserCoolingMOT}%
  \BibitemOpen
  \bibfield  {author} {\bibinfo {author} {\bibfnamefont {L.}~\bibnamefont
  {Anderegg}}, \bibinfo {author} {\bibfnamefont {B.~L.}\ \bibnamefont
  {Augenbraun}}, \bibinfo {author} {\bibfnamefont {Y.}~\bibnamefont {Bao}},
  \bibinfo {author} {\bibfnamefont {S.}~\bibnamefont {Burchesky}}, \bibinfo
  {author} {\bibfnamefont {L.~W.}\ \bibnamefont {Cheuk}}, \bibinfo {author}
  {\bibfnamefont {W.}~\bibnamefont {Ketterle}}, \ and\ \bibinfo {author}
  {\bibfnamefont {J.~M.}\ \bibnamefont {Doyle}},\ }\href {\doibase
  10.1038/s41567-018-0191-z} {\bibfield  {journal} {\bibinfo  {journal} {Nature
  Physics}\ }\textbf {\bibinfo {volume} {14}},\ \bibinfo {pages} {890}
  (\bibinfo {year} {2018})}\BibitemShut {NoStop}%
\bibitem [{\citenamefont {McCarron}(2018)}]{McCarron2018TowardsSlowMolecules}%
  \BibitemOpen
  \bibfield  {author} {\bibinfo {author} {\bibfnamefont {D.~J.}\ \bibnamefont
  {McCarron}},\ }\href {http://stacks.iop.org/1367-2630/20/i=5/a=051001}
  {\bibfield  {journal} {\bibinfo  {journal} {New Journal of Physics}\ }\textbf
  {\bibinfo {volume} {20}},\ \bibinfo {pages} {051001} (\bibinfo {year}
  {2018})}\BibitemShut {NoStop}%
\bibitem [{\citenamefont {Truppe}\ \emph
  {et~al.}(2017{\natexlab{b}})\citenamefont {Truppe}, \citenamefont {Williams},
  \citenamefont {Fitch}, \citenamefont {Hambach}, \citenamefont {Wall},
  \citenamefont {Hinds}, \citenamefont {Sauer},\ and\ \citenamefont
  {Tarbutt}}]{Truppe2017ChirpSlowing}%
  \BibitemOpen
  \bibfield  {author} {\bibinfo {author} {\bibfnamefont {S.}~\bibnamefont
  {Truppe}}, \bibinfo {author} {\bibfnamefont {H.~J.}\ \bibnamefont
  {Williams}}, \bibinfo {author} {\bibfnamefont {N.~J.}\ \bibnamefont {Fitch}},
  \bibinfo {author} {\bibfnamefont {M.}~\bibnamefont {Hambach}}, \bibinfo
  {author} {\bibfnamefont {T.~E.}\ \bibnamefont {Wall}}, \bibinfo {author}
  {\bibfnamefont {E.~A.}\ \bibnamefont {Hinds}}, \bibinfo {author}
  {\bibfnamefont {B.~E.}\ \bibnamefont {Sauer}}, \ and\ \bibinfo {author}
  {\bibfnamefont {M.~R.}\ \bibnamefont {Tarbutt}},\ }\href
  {http://stacks.iop.org/1367-2630/19/i=2/a=022001} {\bibfield  {journal}
  {\bibinfo  {journal} {New Journal of Physics}\ }\textbf {\bibinfo {volume}
  {19}},\ \bibinfo {pages} {022001} (\bibinfo {year}
  {2017}{\natexlab{b}})}\BibitemShut {NoStop}%
\bibitem [{\citenamefont {Barry}\ \emph {et~al.}(2012)\citenamefont {Barry},
  \citenamefont {Shuman}, \citenamefont {Norrgard},\ and\ \citenamefont
  {DeMille}}]{Barry2012RadiationPressureSlowing}%
  \BibitemOpen
  \bibfield  {author} {\bibinfo {author} {\bibfnamefont {J.~F.}\ \bibnamefont
  {Barry}}, \bibinfo {author} {\bibfnamefont {E.~S.}\ \bibnamefont {Shuman}},
  \bibinfo {author} {\bibfnamefont {E.~B.}\ \bibnamefont {Norrgard}}, \ and\
  \bibinfo {author} {\bibfnamefont {D.}~\bibnamefont {DeMille}},\ }\href
  {\doibase 10.1103/PhysRevLett.108.103002} {\bibfield  {journal} {\bibinfo
  {journal} {Phys. Rev. Lett.}\ }\textbf {\bibinfo {volume} {108}},\ \bibinfo
  {pages} {103002} (\bibinfo {year} {2012})}\BibitemShut {NoStop}%
\bibitem [{\citenamefont {Zhelyazkova}\ \emph {et~al.}(2014)\citenamefont
  {Zhelyazkova}, \citenamefont {Cournol}, \citenamefont {Wall}, \citenamefont
  {Matsushima}, \citenamefont {Hudson}, \citenamefont {Hinds}, \citenamefont
  {Tarbutt},\ and\ \citenamefont {Sauer}}]{Zhelyazkova2014CaFCoolingSlowing}%
  \BibitemOpen
  \bibfield  {author} {\bibinfo {author} {\bibfnamefont {V.}~\bibnamefont
  {Zhelyazkova}}, \bibinfo {author} {\bibfnamefont {A.}~\bibnamefont
  {Cournol}}, \bibinfo {author} {\bibfnamefont {T.~E.}\ \bibnamefont {Wall}},
  \bibinfo {author} {\bibfnamefont {A.}~\bibnamefont {Matsushima}}, \bibinfo
  {author} {\bibfnamefont {J.~J.}\ \bibnamefont {Hudson}}, \bibinfo {author}
  {\bibfnamefont {E.~A.}\ \bibnamefont {Hinds}}, \bibinfo {author}
  {\bibfnamefont {M.~R.}\ \bibnamefont {Tarbutt}}, \ and\ \bibinfo {author}
  {\bibfnamefont {B.~E.}\ \bibnamefont {Sauer}},\ }\href {\doibase
  10.1103/PhysRevA.89.053416} {\bibfield  {journal} {\bibinfo  {journal} {Phys.
  Rev. A}\ }\textbf {\bibinfo {volume} {89}},\ \bibinfo {pages} {053416}
  (\bibinfo {year} {2014})}\BibitemShut {NoStop}%
\bibitem [{\citenamefont {Hemmerling}\ \emph {et~al.}(2016)\citenamefont
  {Hemmerling}, \citenamefont {Chae}, \citenamefont {Ravi}, \citenamefont
  {Anderegg}, \citenamefont {Drayna}, \citenamefont {Hutzler}, \citenamefont
  {Collopy}, \citenamefont {Ye}, \citenamefont {Ketterle},\ and\ \citenamefont
  {Doyle}}]{Hemmerling2016CaFWhiteLightSlowing}%
  \BibitemOpen
  \bibfield  {author} {\bibinfo {author} {\bibfnamefont {B.}~\bibnamefont
  {Hemmerling}}, \bibinfo {author} {\bibfnamefont {E.}~\bibnamefont {Chae}},
  \bibinfo {author} {\bibfnamefont {A.}~\bibnamefont {Ravi}}, \bibinfo {author}
  {\bibfnamefont {L.}~\bibnamefont {Anderegg}}, \bibinfo {author}
  {\bibfnamefont {G.~K.}\ \bibnamefont {Drayna}}, \bibinfo {author}
  {\bibfnamefont {N.~R.}\ \bibnamefont {Hutzler}}, \bibinfo {author}
  {\bibfnamefont {A.~L.}\ \bibnamefont {Collopy}}, \bibinfo {author}
  {\bibfnamefont {J.}~\bibnamefont {Ye}}, \bibinfo {author} {\bibfnamefont
  {W.}~\bibnamefont {Ketterle}}, \ and\ \bibinfo {author} {\bibfnamefont
  {J.~M.}\ \bibnamefont {Doyle}},\ }\href
  {http://stacks.iop.org/0953-4075/49/i=17/a=174001} {\bibfield  {journal}
  {\bibinfo  {journal} {Journal of Physics B: Atomic, Molecular and Optical
  Physics}\ }\textbf {\bibinfo {volume} {49}},\ \bibinfo {pages} {174001}
  (\bibinfo {year} {2016})}\BibitemShut {NoStop}%
\bibitem [{\citenamefont {Yeo}\ \emph {et~al.}(2015)\citenamefont {Yeo},
  \citenamefont {Hummon}, \citenamefont {Collopy}, \citenamefont {Yan},
  \citenamefont {Hemmerling}, \citenamefont {Chae}, \citenamefont {Doyle},\
  and\ \citenamefont {Ye}}]{Yeo2015RotationalStateMicrowaveMixing}%
  \BibitemOpen
  \bibfield  {author} {\bibinfo {author} {\bibfnamefont {M.}~\bibnamefont
  {Yeo}}, \bibinfo {author} {\bibfnamefont {M.~T.}\ \bibnamefont {Hummon}},
  \bibinfo {author} {\bibfnamefont {A.~L.}\ \bibnamefont {Collopy}}, \bibinfo
  {author} {\bibfnamefont {B.}~\bibnamefont {Yan}}, \bibinfo {author}
  {\bibfnamefont {B.}~\bibnamefont {Hemmerling}}, \bibinfo {author}
  {\bibfnamefont {E.}~\bibnamefont {Chae}}, \bibinfo {author} {\bibfnamefont
  {J.~M.}\ \bibnamefont {Doyle}}, \ and\ \bibinfo {author} {\bibfnamefont
  {J.}~\bibnamefont {Ye}},\ }\href {\doibase 10.1103/PhysRevLett.114.223003}
  {\bibfield  {journal} {\bibinfo  {journal} {Phys. Rev. Lett.}\ }\textbf
  {\bibinfo {volume} {114}},\ \bibinfo {pages} {223003} (\bibinfo {year}
  {2015})}\BibitemShut {NoStop}%
\bibitem [{\citenamefont {Petzold}\ \emph {et~al.}(2018)\citenamefont
  {Petzold}, \citenamefont {Kaebert}, \citenamefont {Gersema}, \citenamefont
  {Siercke},\ and\ \citenamefont {Ospelkaus}}]{Petzold2018ZeemanSlower}%
  \BibitemOpen
  \bibfield  {author} {\bibinfo {author} {\bibfnamefont {M.}~\bibnamefont
  {Petzold}}, \bibinfo {author} {\bibfnamefont {P.}~\bibnamefont {Kaebert}},
  \bibinfo {author} {\bibfnamefont {P.}~\bibnamefont {Gersema}}, \bibinfo
  {author} {\bibfnamefont {M.}~\bibnamefont {Siercke}}, \ and\ \bibinfo
  {author} {\bibfnamefont {S.}~\bibnamefont {Ospelkaus}},\ }\href@noop {}
  {\bibfield  {journal} {\bibinfo  {journal} {New Journal of Physics}\ }\textbf
  {\bibinfo {volume} {20}},\ \bibinfo {pages} {042001} (\bibinfo {year}
  {2018})}\BibitemShut {NoStop}%
\bibitem [{\citenamefont {Di{R}osa}(2004)}]{DiRosa2004Laser}%
  \BibitemOpen
  \bibfield  {author} {\bibinfo {author} {\bibfnamefont {M.~D.}\ \bibnamefont
  {Di{R}osa}},\ }\href {\doibase 10.1140/epjd/e2004-00167-2} {\bibfield
  {journal} {\bibinfo  {journal} {Eur. Phys. J. D}\ }\textbf {\bibinfo {volume}
  {31}},\ \bibinfo {pages} {395} (\bibinfo {year} {2004})}\BibitemShut
  {NoStop}%
\bibitem [{\citenamefont {Berkeland}\ and\ \citenamefont
  {Boshier}(2002)}]{Berkeland2002Destabilization}%
  \BibitemOpen
  \bibfield  {author} {\bibinfo {author} {\bibfnamefont {D.~J.}\ \bibnamefont
  {Berkeland}}\ and\ \bibinfo {author} {\bibfnamefont {M.~G.}\ \bibnamefont
  {Boshier}},\ }\href {\doibase 10.1103/PhysRevA.65.033413} {\bibfield
  {journal} {\bibinfo  {journal} {Phys. Rev. A}\ }\textbf {\bibinfo {volume}
  {65}},\ \bibinfo {pages} {033413} (\bibinfo {year} {2002})}\BibitemShut
  {NoStop}%
\bibitem [{\citenamefont {Johnson}\ \emph {et~al.}(2015)\citenamefont
  {Johnson}, \citenamefont {Neyenhuis}, \citenamefont {Mizrahi}, \citenamefont
  {Wong-Campos},\ and\ \citenamefont {Monroe}}]{Johnson2015Sensing}%
  \BibitemOpen
  \bibfield  {author} {\bibinfo {author} {\bibfnamefont {K.~G.}\ \bibnamefont
  {Johnson}}, \bibinfo {author} {\bibfnamefont {B.}~\bibnamefont {Neyenhuis}},
  \bibinfo {author} {\bibfnamefont {J.}~\bibnamefont {Mizrahi}}, \bibinfo
  {author} {\bibfnamefont {J.~D.}\ \bibnamefont {Wong-Campos}}, \ and\ \bibinfo
  {author} {\bibfnamefont {C.}~\bibnamefont {Monroe}},\ }\href {\doibase
  10.1103/PhysRevLett.115.213001} {\bibfield  {journal} {\bibinfo  {journal}
  {Phys. Rev. Lett.}\ }\textbf {\bibinfo {volume} {115}},\ \bibinfo {pages}
  {213001} (\bibinfo {year} {2015})}\BibitemShut {NoStop}%
\bibitem [{\citenamefont {Campbell}\ and\ \citenamefont
  {Hamilton}(2017)}]{Campbell2017Rotation}%
  \BibitemOpen
  \bibfield  {author} {\bibinfo {author} {\bibfnamefont {W.~C.}\ \bibnamefont
  {Campbell}}\ and\ \bibinfo {author} {\bibfnamefont {P.}~\bibnamefont
  {Hamilton}},\ }\href {\doibase 10.1088/1361-6455/aa5a8f} {\bibfield
  {journal} {\bibinfo  {journal} {Journal of Physics B: Atomic, Molecular and
  Optical Physics}\ }\textbf {\bibinfo {volume} {50}},\ \bibinfo {pages}
  {064002} (\bibinfo {year} {2017})}\BibitemShut {NoStop}%
\bibitem [{\citenamefont {Garc\'{\i}a-Ripoll}\ \emph
  {et~al.}(2003)\citenamefont {Garc\'{\i}a-Ripoll}, \citenamefont {Zoller},\
  and\ \citenamefont {Cirac}}]{GarciaRipoll2003Speed}%
  \BibitemOpen
  \bibfield  {author} {\bibinfo {author} {\bibfnamefont {J.~J.}\ \bibnamefont
  {Garc\'{\i}a-Ripoll}}, \bibinfo {author} {\bibfnamefont {P.}~\bibnamefont
  {Zoller}}, \ and\ \bibinfo {author} {\bibfnamefont {J.~I.}\ \bibnamefont
  {Cirac}},\ }\href {\doibase 10.1103/PhysRevLett.91.157901} {\bibfield
  {journal} {\bibinfo  {journal} {Phys. Rev. Lett.}\ }\textbf {\bibinfo
  {volume} {91}},\ \bibinfo {pages} {157901} (\bibinfo {year}
  {2003})}\BibitemShut {NoStop}%
\bibitem [{\citenamefont {Duan}(2004)}]{Duan2004Scaling}%
  \BibitemOpen
  \bibfield  {author} {\bibinfo {author} {\bibfnamefont {L.-M.}\ \bibnamefont
  {Duan}},\ }\href {\doibase 10.1103/PhysRevLett.93.100502} {\bibfield
  {journal} {\bibinfo  {journal} {Phys. Rev. Lett.}\ }\textbf {\bibinfo
  {volume} {93}},\ \bibinfo {pages} {100502} (\bibinfo {year}
  {2004})}\BibitemShut {NoStop}%
\bibitem [{\citenamefont {Wong-Campos}\ \emph {et~al.}(2017)\citenamefont
  {Wong-Campos}, \citenamefont {Moses}, \citenamefont {Johnson},\ and\
  \citenamefont {Monroe}}]{WongCampos2017TwoIonGate}%
  \BibitemOpen
  \bibfield  {author} {\bibinfo {author} {\bibfnamefont {J.~D.}\ \bibnamefont
  {Wong-Campos}}, \bibinfo {author} {\bibfnamefont {S.~A.}\ \bibnamefont
  {Moses}}, \bibinfo {author} {\bibfnamefont {K.~G.}\ \bibnamefont {Johnson}},
  \ and\ \bibinfo {author} {\bibfnamefont {C.}~\bibnamefont {Monroe}},\ }\href
  {\doibase 10.1103/PhysRevLett.119.230501} {\bibfield  {journal} {\bibinfo
  {journal} {Phys. Rev. Lett.}\ }\textbf {\bibinfo {volume} {119}},\ \bibinfo
  {pages} {230501} (\bibinfo {year} {2017})}\BibitemShut {NoStop}%
\bibitem [{\citenamefont {Mizrahi}\ \emph {et~al.}(2013)\citenamefont
  {Mizrahi}, \citenamefont {Senko}, \citenamefont {Neyenhuis}, \citenamefont
  {Johnson}, \citenamefont {Campbell}, \citenamefont {Conover},\ and\
  \citenamefont {Monroe}}]{Mizrahi2013Ultrafast}%
  \BibitemOpen
  \bibfield  {author} {\bibinfo {author} {\bibfnamefont {J.}~\bibnamefont
  {Mizrahi}}, \bibinfo {author} {\bibfnamefont {C.}~\bibnamefont {Senko}},
  \bibinfo {author} {\bibfnamefont {B.}~\bibnamefont {Neyenhuis}}, \bibinfo
  {author} {\bibfnamefont {K.~G.}\ \bibnamefont {Johnson}}, \bibinfo {author}
  {\bibfnamefont {W.~C.}\ \bibnamefont {Campbell}}, \bibinfo {author}
  {\bibfnamefont {C.~W.~S.}\ \bibnamefont {Conover}}, \ and\ \bibinfo {author}
  {\bibfnamefont {C.}~\bibnamefont {Monroe}},\ }\href {\doibase
  10.1103/PhysRevLett.110.203001} {\bibfield  {journal} {\bibinfo  {journal}
  {Phys. Rev. Lett.}\ }\textbf {\bibinfo {volume} {110}},\ \bibinfo {pages}
  {203001} (\bibinfo {year} {2013})}\BibitemShut {NoStop}%
\bibitem [{\citenamefont {Jaffe}\ \emph {et~al.}(2018)\citenamefont {Jaffe},
  \citenamefont {Xu}, \citenamefont {Haslinger}, \citenamefont {M\"uller},\
  and\ \citenamefont {Hamilton}}]{Jaffe2018Efficient}%
  \BibitemOpen
  \bibfield  {author} {\bibinfo {author} {\bibfnamefont {M.}~\bibnamefont
  {Jaffe}}, \bibinfo {author} {\bibfnamefont {V.}~\bibnamefont {Xu}}, \bibinfo
  {author} {\bibfnamefont {P.}~\bibnamefont {Haslinger}}, \bibinfo {author}
  {\bibfnamefont {H.}~\bibnamefont {M\"uller}}, \ and\ \bibinfo {author}
  {\bibfnamefont {P.}~\bibnamefont {Hamilton}},\ }\href {\doibase
  10.1103/PhysRevLett.121.040402} {\bibfield  {journal} {\bibinfo  {journal}
  {Phys. Rev. Lett.}\ }\textbf {\bibinfo {volume} {121}},\ \bibinfo {pages}
  {040402} (\bibinfo {year} {2018})}\BibitemShut {NoStop}%
\bibitem [{\citenamefont {Bentley}\ \emph {et~al.}(2015)\citenamefont
  {Bentley}, \citenamefont {Carvalho},\ and\ \citenamefont
  {Hope}}]{Bentley2015IonGate}%
  \BibitemOpen
  \bibfield  {author} {\bibinfo {author} {\bibfnamefont {C.~D.~B.}\
  \bibnamefont {Bentley}}, \bibinfo {author} {\bibfnamefont {A.~R.~R.}\
  \bibnamefont {Carvalho}}, \ and\ \bibinfo {author} {\bibfnamefont {J.~J.}\
  \bibnamefont {Hope}},\ }\href {\doibase 10.1088/1367-2630/17/10/103025}
  {\bibfield  {journal} {\bibinfo  {journal} {New Journal of Physics}\ }\textbf
  {\bibinfo {volume} {17}},\ \bibinfo {pages} {103025} (\bibinfo {year}
  {2015})}\BibitemShut {NoStop}%
\bibitem [{\citenamefont {{Heinrich}}\ \emph {et~al.}(2018)\citenamefont
  {{Heinrich}}, \citenamefont {{Guggemos}}, \citenamefont {{Guevara-Bertsch}},
  \citenamefont {{Hussain}}, \citenamefont {{Roos}},\ and\ \citenamefont
  {{Blatt}}}]{Heinrich2019Ultrafast}%
  \BibitemOpen
  \bibfield  {author} {\bibinfo {author} {\bibfnamefont {D.}~\bibnamefont
  {{Heinrich}}}, \bibinfo {author} {\bibfnamefont {M.}~\bibnamefont
  {{Guggemos}}}, \bibinfo {author} {\bibfnamefont {M.}~\bibnamefont
  {{Guevara-Bertsch}}}, \bibinfo {author} {\bibfnamefont {M.~I.}\ \bibnamefont
  {{Hussain}}}, \bibinfo {author} {\bibfnamefont {C.~F.}\ \bibnamefont
  {{Roos}}}, \ and\ \bibinfo {author} {\bibfnamefont {R.}~\bibnamefont
  {{Blatt}}},\ }\href@noop {} {\bibfield  {journal} {\bibinfo  {journal} {arXiv
  e-prints}\ ,\ \bibinfo {eid} {arXiv:1812.08537}} (\bibinfo {year} {2018})},\
  \Eprint {http://arxiv.org/abs/1812.08537} {arXiv:1812.08537 [quant-ph]}
  \BibitemShut {NoStop}%
\bibitem [{\citenamefont {Kazantsev}(1974)}]{Kazantsev1974TheAcceleration}%
  \BibitemOpen
  \bibfield  {author} {\bibinfo {author} {\bibfnamefont {A.~P.}\ \bibnamefont
  {Kazantsev}},\ }\href@noop {} {\bibfield  {journal} {\bibinfo  {journal}
  {Sov. Phys.-JETP}\ }\textbf {\bibinfo {volume} {39}},\ \bibinfo {pages} {784}
  (\bibinfo {year} {1974})}\BibitemShut {NoStop}%
\bibitem [{\citenamefont {Jayich}\ \emph {et~al.}(2014)\citenamefont {Jayich},
  \citenamefont {Vutha}, \citenamefont {Hummon}, \citenamefont {Porto},\ and\
  \citenamefont {Campbell}}]{Jayich2014theoryProposal}%
  \BibitemOpen
  \bibfield  {author} {\bibinfo {author} {\bibfnamefont {A.~M.}\ \bibnamefont
  {Jayich}}, \bibinfo {author} {\bibfnamefont {A.~C.}\ \bibnamefont {Vutha}},
  \bibinfo {author} {\bibfnamefont {M.~T.}\ \bibnamefont {Hummon}}, \bibinfo
  {author} {\bibfnamefont {J.~V.}\ \bibnamefont {Porto}}, \ and\ \bibinfo
  {author} {\bibfnamefont {W.~C.}\ \bibnamefont {Campbell}},\ }\href {\doibase
  10.1103/PhysRevA.89.023425} {\bibfield  {journal} {\bibinfo  {journal} {Phys.
  Rev. A}\ }\textbf {\bibinfo {volume} {89}},\ \bibinfo {pages} {023425}
  (\bibinfo {year} {2014})}\BibitemShut {NoStop}%
\bibitem [{\citenamefont {Vo\u{\i}tsekhovich}\ \emph
  {et~al.}(1994)\citenamefont {Vo\u{\i}tsekhovich}, \citenamefont
  {Danile\u{\i}ko}, \citenamefont {Negri\u{\i}ko}, \citenamefont {Romanenko},\
  and\ \citenamefont {Yatsenko}}]{Voitsekhovich1994Observation}%
  \BibitemOpen
  \bibfield  {author} {\bibinfo {author} {\bibfnamefont {V.~S.}\ \bibnamefont
  {Vo\u{\i}tsekhovich}}, \bibinfo {author} {\bibfnamefont {M.~V.}\ \bibnamefont
  {Danile\u{\i}ko}}, \bibinfo {author} {\bibfnamefont {A.~M.}\ \bibnamefont
  {Negri\u{\i}ko}}, \bibinfo {author} {\bibfnamefont {V.~I.}\ \bibnamefont
  {Romanenko}}, \ and\ \bibinfo {author} {\bibfnamefont {L.~P.}\ \bibnamefont
  {Yatsenko}},\ }\href@noop {} {\bibfield  {journal} {\bibinfo  {journal} {JETP
  Lett.}\ }\textbf {\bibinfo {volume} {59}},\ \bibinfo {pages} {408} (\bibinfo
  {year} {1994})}\BibitemShut {NoStop}%
\bibitem [{\citenamefont {N\"olle}\ \emph {et~al.}(1996)\citenamefont
  {N\"olle}, \citenamefont {N\"olle}, \citenamefont {Schmand},\ and\
  \citenamefont {Andr\"a}}]{Nolle1996AtomicBeam}%
  \BibitemOpen
  \bibfield  {author} {\bibinfo {author} {\bibfnamefont {B.}~\bibnamefont
  {N\"olle}}, \bibinfo {author} {\bibfnamefont {H.}~\bibnamefont {N\"olle}},
  \bibinfo {author} {\bibfnamefont {J.}~\bibnamefont {Schmand}}, \ and\
  \bibinfo {author} {\bibfnamefont {H.~J.}\ \bibnamefont {Andr\"a}},\
  }\href@noop {} {\bibfield  {journal} {\bibinfo  {journal} {Europhys. Lett.}\
  }\textbf {\bibinfo {volume} {33}},\ \bibinfo {pages} {261} (\bibinfo {year}
  {1996})}\BibitemShut {NoStop}%
\bibitem [{\citenamefont {Goepfert}\ \emph {et~al.}(1997)\citenamefont
  {Goepfert}, \citenamefont {Bloch}, \citenamefont {Haubrich}, \citenamefont
  {Lison}, \citenamefont {Sch\"utze}, \citenamefont {Wynands},\ and\
  \citenamefont {Meschede}}]{Goepfert1997Stimulated}%
  \BibitemOpen
  \bibfield  {author} {\bibinfo {author} {\bibfnamefont {A.}~\bibnamefont
  {Goepfert}}, \bibinfo {author} {\bibfnamefont {I.}~\bibnamefont {Bloch}},
  \bibinfo {author} {\bibfnamefont {D.}~\bibnamefont {Haubrich}}, \bibinfo
  {author} {\bibfnamefont {F.}~\bibnamefont {Lison}}, \bibinfo {author}
  {\bibfnamefont {R.}~\bibnamefont {Sch\"utze}}, \bibinfo {author}
  {\bibfnamefont {R.}~\bibnamefont {Wynands}}, \ and\ \bibinfo {author}
  {\bibfnamefont {D.}~\bibnamefont {Meschede}},\ }\href@noop {} {\bibfield
  {journal} {\bibinfo  {journal} {Phys. Rev. A}\ }\textbf {\bibinfo {volume}
  {56}},\ \bibinfo {pages} {R3354} (\bibinfo {year} {1997})}\BibitemShut
  {NoStop}%
\bibitem [{\citenamefont {Kozyryev}\ \emph
  {et~al.}(2018{\natexlab{a}})\citenamefont {Kozyryev}, \citenamefont {Baum},
  \citenamefont {Matsuda},\ and\ \citenamefont {Doyle}}]{Kozyryev2016Proposal}%
  \BibitemOpen
  \bibfield  {author} {\bibinfo {author} {\bibfnamefont {I.}~\bibnamefont
  {Kozyryev}}, \bibinfo {author} {\bibfnamefont {L.}~\bibnamefont {Baum}},
  \bibinfo {author} {\bibfnamefont {K.}~\bibnamefont {Matsuda}}, \ and\
  \bibinfo {author} {\bibfnamefont {J.~M.}\ \bibnamefont {Doyle}},\ }\href@noop
  {} {\bibfield  {journal} {\bibinfo  {journal} {ChemPhysChem}\ }\textbf
  {\bibinfo {volume} {17}},\ \bibinfo {pages} {3641} (\bibinfo {year}
  {2018}{\natexlab{a}})}\BibitemShut {NoStop}%
\bibitem [{\citenamefont {{Denis}}\ \emph {et~al.}(2019)\citenamefont
  {{Denis}}, \citenamefont {{Haase}}, \citenamefont {{Timmermans}},
  \citenamefont {{Eliav}}, \citenamefont {{Hutzler}},\ and\ \citenamefont
  {{Borschevsky}}}]{Denis2019Enhancement}%
  \BibitemOpen
  \bibfield  {author} {\bibinfo {author} {\bibfnamefont {M.}~\bibnamefont
  {{Denis}}}, \bibinfo {author} {\bibfnamefont {P.~A.~B.}\ \bibnamefont
  {{Haase}}}, \bibinfo {author} {\bibfnamefont {R.~G.~E.}\ \bibnamefont
  {{Timmermans}}}, \bibinfo {author} {\bibfnamefont {E.}~\bibnamefont
  {{Eliav}}}, \bibinfo {author} {\bibfnamefont {N.~R.}\ \bibnamefont
  {{Hutzler}}}, \ and\ \bibinfo {author} {\bibfnamefont {A.}~\bibnamefont
  {{Borschevsky}}},\ }\href@noop {} {\bibfield  {journal} {\bibinfo  {journal}
  {arXiv e-prints}\ ,\ \bibinfo {eid} {arXiv:1901.02265}} (\bibinfo {year}
  {2019})},\ \Eprint {http://arxiv.org/abs/1901.02265} {arXiv:1901.02265
  [physics.atom-ph]} \BibitemShut {NoStop}%
\bibitem [{\citenamefont {Lim}\ \emph {et~al.}(2018)\citenamefont {Lim},
  \citenamefont {Almond}, \citenamefont {Trigatzis}, \citenamefont {Devlin},
  \citenamefont {Fitch}, \citenamefont {Sauer}, \citenamefont {Tarbutt},\ and\
  \citenamefont {Hinds}}]{Lim2018YbF}%
  \BibitemOpen
  \bibfield  {author} {\bibinfo {author} {\bibfnamefont {J.}~\bibnamefont
  {Lim}}, \bibinfo {author} {\bibfnamefont {J.~R.}\ \bibnamefont {Almond}},
  \bibinfo {author} {\bibfnamefont {M.~A.}\ \bibnamefont {Trigatzis}}, \bibinfo
  {author} {\bibfnamefont {J.~A.}\ \bibnamefont {Devlin}}, \bibinfo {author}
  {\bibfnamefont {N.~J.}\ \bibnamefont {Fitch}}, \bibinfo {author}
  {\bibfnamefont {B.~E.}\ \bibnamefont {Sauer}}, \bibinfo {author}
  {\bibfnamefont {M.~R.}\ \bibnamefont {Tarbutt}}, \ and\ \bibinfo {author}
  {\bibfnamefont {E.~A.}\ \bibnamefont {Hinds}},\ }\href {\doibase
  10.1103/PhysRevLett.120.123201} {\bibfield  {journal} {\bibinfo  {journal}
  {Phys. Rev. Lett.}\ }\textbf {\bibinfo {volume} {120}},\ \bibinfo {pages}
  {123201} (\bibinfo {year} {2018})}\BibitemShut {NoStop}%
\bibitem [{\citenamefont {Hunter}\ \emph {et~al.}(2012)\citenamefont {Hunter},
  \citenamefont {Peck}, \citenamefont {Greenspon}, \citenamefont {Alam},\ and\
  \citenamefont {DeMille}}]{Hunter2012TlF}%
  \BibitemOpen
  \bibfield  {author} {\bibinfo {author} {\bibfnamefont {L.~R.}\ \bibnamefont
  {Hunter}}, \bibinfo {author} {\bibfnamefont {S.~K.}\ \bibnamefont {Peck}},
  \bibinfo {author} {\bibfnamefont {A.~S.}\ \bibnamefont {Greenspon}}, \bibinfo
  {author} {\bibfnamefont {S.~S.}\ \bibnamefont {Alam}}, \ and\ \bibinfo
  {author} {\bibfnamefont {D.}~\bibnamefont {DeMille}},\ }\href {\doibase
  10.1103/PhysRevA.85.012511} {\bibfield  {journal} {\bibinfo  {journal} {Phys.
  Rev. A}\ }\textbf {\bibinfo {volume} {85}},\ \bibinfo {pages} {012511}
  (\bibinfo {year} {2012})}\BibitemShut {NoStop}%
\bibitem [{\citenamefont {Ip}\ \emph {et~al.}(2018)\citenamefont {Ip},
  \citenamefont {Ransford}, \citenamefont {Jayich}, \citenamefont {Long},
  \citenamefont {Roman},\ and\ \citenamefont {Campbell}}]{Ip2018Phonon}%
  \BibitemOpen
  \bibfield  {author} {\bibinfo {author} {\bibfnamefont {M.}~\bibnamefont
  {Ip}}, \bibinfo {author} {\bibfnamefont {A.}~\bibnamefont {Ransford}},
  \bibinfo {author} {\bibfnamefont {A.~M.}\ \bibnamefont {Jayich}}, \bibinfo
  {author} {\bibfnamefont {X.}~\bibnamefont {Long}}, \bibinfo {author}
  {\bibfnamefont {C.}~\bibnamefont {Roman}}, \ and\ \bibinfo {author}
  {\bibfnamefont {W.~C.}\ \bibnamefont {Campbell}},\ }\href@noop {} {\bibfield
  {journal} {\bibinfo  {journal} {Phys. Rev. Lett.}\ }\textbf {\bibinfo
  {volume} {121}},\ \bibinfo {pages} {043201} (\bibinfo {year}
  {2018})}\BibitemShut {NoStop}%
\bibitem [{\citenamefont {Jacobs}()}]{Verne}%
  \BibitemOpen
  \bibfield  {author} {\bibinfo {author} {\bibfnamefont {V.~L.}\ \bibnamefont
  {Jacobs}},\ }\href@noop {} {}\bibinfo {howpublished} {manuscript in
  preparation}\BibitemShut {NoStop}%
\bibitem [{Sup()}]{Supplementary}%
  \BibitemOpen
  \href@noop {} {}\bibinfo {note} {See Supplemental Material at [URL will be
  inserted by publisher] for details on comb tooth effects and TOF
  measurements, which includes Refs.
  \cite{Felinto2003PulseCoherentAccumulation, Llinova2011CombCoolingTheory,
  Aumiler2012CombSimultaneousCooling}}\BibitemShut {NoStop}%
\bibitem [{\citenamefont {Kozyryev}\ \emph
  {et~al.}(2018{\natexlab{b}})\citenamefont {Kozyryev}, \citenamefont {Baum},
  \citenamefont {Aldridge}, \citenamefont {Yu}, \citenamefont {Eyler},\ and\
  \citenamefont {Doyle}}]{Kozyryev2018BichromaticForce}%
  \BibitemOpen
  \bibfield  {author} {\bibinfo {author} {\bibfnamefont {I.}~\bibnamefont
  {Kozyryev}}, \bibinfo {author} {\bibfnamefont {L.}~\bibnamefont {Baum}},
  \bibinfo {author} {\bibfnamefont {L.}~\bibnamefont {Aldridge}}, \bibinfo
  {author} {\bibfnamefont {P.}~\bibnamefont {Yu}}, \bibinfo {author}
  {\bibfnamefont {E.~E.}\ \bibnamefont {Eyler}}, \ and\ \bibinfo {author}
  {\bibfnamefont {J.~M.}\ \bibnamefont {Doyle}},\ }\href {\doibase
  10.1103/PhysRevLett.120.063205} {\bibfield  {journal} {\bibinfo  {journal}
  {Phys. Rev. Lett.}\ }\textbf {\bibinfo {volume} {120}},\ \bibinfo {pages}
  {063205} (\bibinfo {year} {2018}{\natexlab{b}})}\BibitemShut {NoStop}%
\bibitem [{\citenamefont {Kozyryev}()}]{Kozyryev2018PrivateComminucation}%
  \BibitemOpen
  \bibfield  {author} {\bibinfo {author} {\bibfnamefont {I.}~\bibnamefont
  {Kozyryev}},\ }\href@noop {} {}\bibinfo {howpublished} {private
  communication}\BibitemShut {NoStop}%
\bibitem [{\citenamefont {Galica}\ \emph {et~al.}(2013)\citenamefont {Galica},
  \citenamefont {Aldridge},\ and\ \citenamefont
  {Eyler}}]{Galica2013four_color}%
  \BibitemOpen
  \bibfield  {author} {\bibinfo {author} {\bibfnamefont {S.~E.}\ \bibnamefont
  {Galica}}, \bibinfo {author} {\bibfnamefont {L.}~\bibnamefont {Aldridge}}, \
  and\ \bibinfo {author} {\bibfnamefont {E.~E.}\ \bibnamefont {Eyler}},\ }\href
  {\doibase 10.1103/PhysRevA.88.043418} {\bibfield  {journal} {\bibinfo
  {journal} {Phys. Rev. A}\ }\textbf {\bibinfo {volume} {88}},\ \bibinfo
  {pages} {043418} (\bibinfo {year} {2013})}\BibitemShut {NoStop}%
\bibitem [{\citenamefont {Galica}\ \emph {et~al.}(2018)\citenamefont {Galica},
  \citenamefont {Aldridge}, \citenamefont {McCarron}, \citenamefont {Eyler},\
  and\ \citenamefont {Gould}}]{Galica2018BichromaticDeflection}%
  \BibitemOpen
  \bibfield  {author} {\bibinfo {author} {\bibfnamefont {S.~E.}\ \bibnamefont
  {Galica}}, \bibinfo {author} {\bibfnamefont {L.}~\bibnamefont {Aldridge}},
  \bibinfo {author} {\bibfnamefont {D.~J.}\ \bibnamefont {McCarron}}, \bibinfo
  {author} {\bibfnamefont {E.~E.}\ \bibnamefont {Eyler}}, \ and\ \bibinfo
  {author} {\bibfnamefont {P.~L.}\ \bibnamefont {Gould}},\ }\href {\doibase
  10.1103/PhysRevA.98.023408} {\bibfield  {journal} {\bibinfo  {journal} {Phys.
  Rev. A}\ }\textbf {\bibinfo {volume} {98}},\ \bibinfo {pages} {023408}
  (\bibinfo {year} {2018})}\BibitemShut {NoStop}%
\bibitem [{\citenamefont {Kozyryev}\ \emph {et~al.}(2016)\citenamefont
  {Kozyryev}, \citenamefont {Baum}, \citenamefont {Matsuda}, \citenamefont
  {Hemmerling},\ and\ \citenamefont
  {Doyle}}]{Kozyryev2016PolyatomicRadiativeForce}%
  \BibitemOpen
  \bibfield  {author} {\bibinfo {author} {\bibfnamefont {I.}~\bibnamefont
  {Kozyryev}}, \bibinfo {author} {\bibfnamefont {L.}~\bibnamefont {Baum}},
  \bibinfo {author} {\bibfnamefont {K.}~\bibnamefont {Matsuda}}, \bibinfo
  {author} {\bibfnamefont {B.}~\bibnamefont {Hemmerling}}, \ and\ \bibinfo
  {author} {\bibfnamefont {J.~M.}\ \bibnamefont {Doyle}},\ }\href {\doibase
  10.1088/0953-4075/49/13/134002} {\bibfield  {journal} {\bibinfo  {journal}
  {Journal of Physics B: Atomic, Molecular and Optical Physics}\ }\textbf
  {\bibinfo {volume} {49}},\ \bibinfo {pages} {134002} (\bibinfo {year}
  {2016})}\BibitemShut {NoStop}%
\bibitem [{\citenamefont {Shuman}\ \emph {et~al.}(2009)\citenamefont {Shuman},
  \citenamefont {Barry}, \citenamefont {Glenn},\ and\ \citenamefont
  {DeMille}}]{Shuman2009DiatomicRadiativeForce}%
  \BibitemOpen
  \bibfield  {author} {\bibinfo {author} {\bibfnamefont {E.~S.}\ \bibnamefont
  {Shuman}}, \bibinfo {author} {\bibfnamefont {J.~F.}\ \bibnamefont {Barry}},
  \bibinfo {author} {\bibfnamefont {D.~R.}\ \bibnamefont {Glenn}}, \ and\
  \bibinfo {author} {\bibfnamefont {D.}~\bibnamefont {DeMille}},\ }\href
  {\doibase 10.1103/PhysRevLett.103.223001} {\bibfield  {journal} {\bibinfo
  {journal} {Phys. Rev. Lett.}\ }\textbf {\bibinfo {volume} {103}},\ \bibinfo
  {pages} {223001} (\bibinfo {year} {2009})}\BibitemShut {NoStop}%
\bibitem [{\citenamefont {Nakhate}\ \emph {et~al.}(2019)\citenamefont
  {Nakhate}, \citenamefont {Steimle}, \citenamefont {Pilgram},\ and\
  \citenamefont {Hutzler}}]{Nakhate2018YbOHSpectrum}%
  \BibitemOpen
  \bibfield  {author} {\bibinfo {author} {\bibfnamefont {S.}~\bibnamefont
  {Nakhate}}, \bibinfo {author} {\bibfnamefont {T.~C.}\ \bibnamefont
  {Steimle}}, \bibinfo {author} {\bibfnamefont {N.~H.}\ \bibnamefont
  {Pilgram}}, \ and\ \bibinfo {author} {\bibfnamefont {N.~R.}\ \bibnamefont
  {Hutzler}},\ }\href {\doibase https://doi.org/10.1016/j.cplett.2018.11.030}
  {\bibfield  {journal} {\bibinfo  {journal} {Chemical Physics Letters}\
  }\textbf {\bibinfo {volume} {715}},\ \bibinfo {pages} {105 } (\bibinfo {year}
  {2019})}\BibitemShut {NoStop}%
\bibitem [{\citenamefont {Felinto}\ \emph {et~al.}(2003)\citenamefont
  {Felinto}, \citenamefont {Bosco}, \citenamefont {Acioli},\ and\ \citenamefont
  {Vianna}}]{Felinto2003PulseCoherentAccumulation}%
  \BibitemOpen
  \bibfield  {author} {\bibinfo {author} {\bibfnamefont {D.}~\bibnamefont
  {Felinto}}, \bibinfo {author} {\bibfnamefont {C.}~\bibnamefont {Bosco}},
  \bibinfo {author} {\bibfnamefont {L.}~\bibnamefont {Acioli}}, \ and\ \bibinfo
  {author} {\bibfnamefont {S.}~\bibnamefont {Vianna}},\ }\href {\doibase
  https://doi.org/10.1016/S0030-4018(02)02230-7} {\bibfield  {journal}
  {\bibinfo  {journal} {Optics Communications}\ }\textbf {\bibinfo {volume}
  {215}},\ \bibinfo {pages} {69 } (\bibinfo {year} {2003})}\BibitemShut
  {NoStop}%
\bibitem [{\citenamefont {Ilinova}\ \emph {et~al.}(2011)\citenamefont
  {Ilinova}, \citenamefont {Ahmad},\ and\ \citenamefont
  {Derevianko}}]{Llinova2011CombCoolingTheory}%
  \BibitemOpen
  \bibfield  {author} {\bibinfo {author} {\bibfnamefont {E.}~\bibnamefont
  {Ilinova}}, \bibinfo {author} {\bibfnamefont {M.}~\bibnamefont {Ahmad}}, \
  and\ \bibinfo {author} {\bibfnamefont {A.}~\bibnamefont {Derevianko}},\
  }\href {\doibase 10.1103/PhysRevA.84.033421} {\bibfield  {journal} {\bibinfo
  {journal} {Phys. Rev. A}\ }\textbf {\bibinfo {volume} {84}},\ \bibinfo
  {pages} {033421} (\bibinfo {year} {2011})}\BibitemShut {NoStop}%
\bibitem [{\citenamefont {Aumiler}\ and\ \citenamefont
  {Ban}(2012)}]{Aumiler2012CombSimultaneousCooling}%
  \BibitemOpen
  \bibfield  {author} {\bibinfo {author} {\bibfnamefont {D.}~\bibnamefont
  {Aumiler}}\ and\ \bibinfo {author} {\bibfnamefont {T.}~\bibnamefont {Ban}},\
  }\href {\doibase 10.1103/PhysRevA.85.063412} {\bibfield  {journal} {\bibinfo
  {journal} {Phys. Rev. A}\ }\textbf {\bibinfo {volume} {85}},\ \bibinfo
  {pages} {063412} (\bibinfo {year} {2012})}\BibitemShut {NoStop}%
\end{thebibliography}%
\end{document}